\documentclass[universe,article,accept,pdftex,moreauthors]{Definitions/mdpi}

\firstpage{1} 
\pubvolume{1}
\issuenum{1}
\articlenumber{0}
\pubyear{2025}
\copyrightyear{2025}
\externaleditor{Firstname Lastname} 
\datereceived{31 October 2025} 
\daterevised{28 November 2025} 
\dateaccepted{4 December 2025} 
\datepublished{ } 
\hreflink{https://doi.org/} 

\Title{Exploring the Role of Vector Potential and Plasma-$\beta$ in Jet Formation from Magnetized Accretion Flows}

\TitleCitation{Exploring the Role of Vector Potential and Plasma-$\beta$ in Jet Formation from Magnetized Accretion Flows}

\Author{Ishika 
 Palit $^{1,}$*\orcidA{}, Miles Angelo Paloma Sodejana
 $^{2}$\orcidB{} and Hsiang-Yi Karen Yang $^{1,3}$\orcidC{}}

\AuthorNames{Ishika Palit, Miles Angelo P. Sodejana and Hsiang-Yi Karen Yang}

\AuthorCitation{Palit, 
 I.; Sodejana, M.A.P.; Yang, H.-Y.K.}

\address{%
$^{1}$ \quad Department of Physics, Institute of Astronomy 
 National Tsing Hua University, Hsinchu 30013, Taiwan; hyang@phys.nthu.edu.tw 
\\
$^{2}$ \quad Department of Physics, University of Southern Mindanao--Main, 
 Kabacan 9407, Cotabato, 
 Philippines; mapsodejana@usm.edu.ph\\
$^{3}$ \quad Physics Division, National Center for Theoretical Sciences, Taipei 10617, Taiwan}

\corres{Correspondence: v01853@phys.nthu.edu.tw}

\abstract{
In this work, we investigate how the choice of initial vector potential and plasma parameters influences the development of accretion columns and jet formation in magnetized accretion flows. Using general relativistic magnetohydrodynamic simulations, we explore two different configurations of the vector potential \( A_\phi \) and three plasma beta values \mbox{(\( \beta = 50, 100, 500 \))}. We analyze how variations in the poloidal magnetic field strength and plasma magnetization affect magnetic flux accumulation near the black hole and the subsequent growth of the accretion column. Our results highlight the dependence of jet launching efficiency and accretion dynamics on the initial magnetic field topology and plasma beta, offering insight into the conditions that favor magnetically arrested disk or standard and normal \mbox{evolution states}.}


\begin{document}

\section{Introduction}
Accretion onto compact objects, such as black holes, neutron stars, and white dwarfs, is one of the most energetic and fundamental processes in astrophysics. In black hole systems in particular, accretion often gives rise to powerful outflows in the form of winds and relativistic jets, which play a central role in shaping their surroundings across cosmic scales (see \cite{McKinney2013, bardeen1975, lodato2006} and references therein). From X-ray binaries to active galactic nuclei (AGN), these systems emit across the entire electromagnetic spectrum and provide key observational probes of strong-field gravity, magnetohydrodynamics, and plasma physics in \mbox{extreme environments}.

Theoretical and computational advances over the past few decades have significantly improved our understanding of the coupling between magnetic fields, plasma dynamics, and relativistic jet formation near black holes (see \cite{Tchekhovskoy2011-4, Chatterjee2022, Salas2024} and references therein). General relativistic magnetohydrodynamic (GRMHD) simulations, in particular, have proven indispensable in studying how magnetic flux accumulation and turbulence govern accretion behavior and jet launching. Such simulations have revealed two broad categories of accretion states: the standard and normal evolution (SANE) regime, where magnetic fields remain dynamically subdominant, and the magnetically arrested disk (MAD) regime, in which strong poloidal fields inhibit accretion and drive powerful jets \cite{Tchekhovskoy2011-4}. The balance between these states depends sensitively on the initial magnetic field configuration, plasma magnetization, and accretion rate.

The Event Horizon Telescope (EHT) observations of M87* and Sgr~A* have provided compelling evidence that some supermassive black holes operate in magnetically dominated, MAD-like conditions [Event Horizon Telescope Collaboration \cite{eht2024}, as also discussed in the references cited therein]. Synthetic images derived from  GRMHD simulations successfully reproduce the observed emission morphology, further underscoring the need for detailed modeling of magnetic field evolution in accretion flows \cite{chael2019, chan2025}. However, despite rapid progress, key questions remain regarding the dependence of jet launching and magnetic flux accumulation on initial magnetic topology and plasma parameters \cite{Palit2025}. In particular, the role of the initial vector potential---defining the structure and strength of the magnetic field threading the accretion torus---remains only partially understood.

In this work, we explore how the choice of initial vector potential \( A_\phi \) and the plasma beta parameter (\( \beta = P_{\mathrm{gas}} / P_{\mathrm{mag}} \)) influence the development of the accretion column and the onset of magnetically dominated behavior. By varying the magnetic field strength (through different values of plasma beta) and testing two distinct prescriptions for \( A_\phi \), we study the resulting evolution of magnetic flux accumulation and jet formation around the black hole. These simulations allow us to identify how the poloidal magnetic field configuration and plasma magnetization regulate the efficiency of jet launching and the overall \mbox{accretion dynamics}.

This approach provides an understanding of how the interplay between plasma beta and vector potential shapes the magnetic field geometry near the event horizon. Such insights are essential for interpreting high-resolution observations from the EHT and future instruments, and for constraining theoretical models of accretion-powered jets in both stellar-mass and supermassive black holes.

The paper is organized as follows. Section~\ref{sec.2} describes the simulation setup, including the initial conditions, choice of vector potential, and plasma parameters. Section~\ref{sec.3} presents the resulting accretion dynamics and jet structure for each case. Section~\ref{sec.4} discusses the implications of our findings for the magnetization and variability of accretion flows. 

\section{Numerical Setup} \label{sec.2}

We
 employ the GRMHD code HARM 
 \cite{Gammie2003, Noble2006, Sapountzis2019} for our numerical scheme with the Kerr spacetime as the background. The code is a conservative, shock-capturing scheme that solves the continuity, energy–momentum conservation, and induction equations \mbox{in GRMHD} 
\begin{align}
    \nabla_\mu (\rho u^\mu) &= 0, \label{continuity} \\ \nabla_\mu (T^{\mu\nu}) &= 0, \label{energymomentum}\\\nabla_\mu (u^\nu b^\mu - u^\mu b^\nu) &= 0 \label{induction},
\end{align}
where $\rho$, $u^\mu$, and $b^\mu$ denote the gas density, gas four-velocity, and magnetic four-vector, while $T^{\mu\nu} = T^{\mu\nu}_\mathrm{gas} + T^{\mu\nu}_\mathrm{EM}$ is the stress-energy tensor with both gas and electromagnetic components. Here, 
\begin{align}
    T^{\mu\nu}_\mathrm{gas} &= (\rho + u + p)u^\mu u^\nu + pg^{\mu\nu}, \label{Tgas}\end{align} 
and 
    \begin{align}T^{\mu\nu}_\mathrm{EM} &= b^2 u^\mu u^\nu + \frac{1}{2}b^2 g^{\mu\nu}- b^\mu b^\nu, \ \ \ b^\mu = u_\nu^* F^{\mu\nu}, \label{Tem}  
\end{align}where $u, \ p$, and $^* F^{\mu\nu}$ denote the internal energy, pressure, and the dual of the Faraday (electromagnetic) tensor. We adopt geometrized units with $G = c = M = 1$, where $M$ is the black hole mass, such that the length and time in code units are given by $r_g = GM/c^2$ and $t_g = GM/c^3$.

We evolve the system in modified Kerr–Schild coordinates, with the initial Fishbone–Moncrief torus \cite{FM76} specified in Boyer–Lindquist coordinates and transformed to Kerr–Schild coordinates within the code (see \cite{Sapountzis2019} \& references therein). The inner radius of the torus is at $r_{\mathrm{in}} = 6\,r_g$ and its pressure maximum is located at $r_{\mathrm{max}} = 13\,r_g$. We adopt an ideal-gas equation of state with adiabatic index $\gamma = 4/3$, appropriate for relativistically hot plasma.

The computational domain extends from $r_{\mathrm{min}} = 1.2\,r_g$ to $r_{\mathrm{max}} = 10^3\,r_g$, fully covering the polar angles $\theta \in [0,\pi]$ and azimuthal angles $\phi \in [0, 2\pi]$. 
We employ logarithmic spacing in the radial direction, a modified-sine coordinate for $\theta$ that concentrates resolution toward the equatorial plane, and uniform spacing in $\phi$. The numerical resolution is $384 \times 256 \times 128$ in $(r,\theta,\phi)$.

We apply outflow (zero-gradient) boundary conditions at both the inner and outer radial boundaries. The inner boundary is placed inside the black-hole event horizon for smooth flow. The azimuthal ($\phi$) direction employs periodic boundary conditions.


The black hole magnetosphere is initially threaded by large-scale poloidal magnetic field loops, introduced through the azimuthal component of the vector potential $A_\phi$, which determines the field topology within the accretion torus. We consider two distinct configurations of the vector potential, denoted as $A_{\phi}^{(1)}$ and $A_{\phi}^{(2)}$, corresponding to black hole spin parameters $a = 0.9$ and $a = 0.935$, respectively. Both configurations evolve toward a MAD state \cite{Igor2003, Narayan2003, Tchekhovskoy2011-4, Chatterjee2022, James2022, Salas2024}. The vector potentials are defined as follows:
\begin{align}
A_{\phi}^{(1)}(r, \theta) &= r^5 \left( \frac{\rho_\mathrm{avg}}{\rho_\mathrm{max}} \right) - 0.2, \label{Aphi1} \\
A_{\phi}^{(2)}(r, \theta) &= \frac{\rho}{\rho_\mathrm{max}} \left( \frac{r}{r_\mathrm{in}} \right)^3 \sin^3\theta \, \exp\left(-\frac{r}{400}\right) - 0.2. \label{Aphi2}
\end{align}
Throughout this work, we refer to these configurations as $A_{\phi}^{(1)}$ (Equation~(\ref{Aphi1})) and $A_{\phi}^{(2)}$ (Equation~(\ref{Aphi2})), respectively. Here, $\rho$ and $\rho_\mathrm{avg}$ denote the local and shell-averaged rest-mass density, respectively, while $\rho_\mathrm{max}$ is the maximum density within the initial torus.

The configuration $A_{\phi}^{(1)}$ represents a large-scale, coherent poloidal field with strong initial flux near the inner disk. This setup promotes rapid angular momentum transport and efficient advection of magnetic flux toward the black hole, leading to faster evolution of the accretion torus and earlier jet launching. This setup is relevant for systems such as collapsar central engines, where magnetic flux amplification and advection occur quickly during core collapse \cite{James2022}. In contrast, $A_{\phi}^{(2)}$ corresponds to a more compact, disk-centered field topology with weaker initial poloidal flux. The accretion flow in this configuration evolves more gradually, allowing a slower buildup toward a MAD state and smoother, more ordered jet structures. This setup is relevant for post-merger remnants or active galactic nuclei, where magnetic flux is already concentrated near the inner disk \cite{Chatterjee2022}. These two configurations thus allow us to explore how initial magnetic geometry and flux distribution influence the early stages of accretion and jet formation.

The magnetic field strength and degree of magnetization are characterized by the plasma beta $\beta$ and the magnetization parameter $\sigma$:
\begin{align}
\beta = \frac{p_{\mathrm{gas}}}{p_{\mathrm{mag}}}, \hspace{5mm} \sigma = \frac{b^2}{\rho}, \label{betasigma}
\end{align}
where $p_{\mathrm{gas}}$ is the gas pressure, $p_{\mathrm{mag}} = b^2/2$ is the magnetic pressure, and $b^2 = b_\mu b^\mu$ is the magnetic field strength measured in the comoving frame.

We have simulated 5 different models. The first vector potential configuration, $A_{\phi}^{(1)}$, is explored with plasma beta values $\beta = 50$ and $100$, while the second configuration, $A_{\phi}^{(2)}$, is studied with $\beta = 50$, $100$, and $500$.


\section{Results}\label{sec.3}

In this section, we present the results of our simulations for five different models. In Section~\ref{3.2}, we show two-dimensional (2D) maps of key physical quantities, including density, magnetic field strength, and related variables, along both poloidal and equatorial planes. In Section~\ref{3.3}, we examine the temporal evolution of these quantities, illustrating how the accretion flow, magnetic flux, and torus structure develop over time under different magnetic field configurations and plasma parameters.

\subsection{Spatial Structure of the Accretion Flow}
\label{3.2}
\begin{figure}[H]
    \includegraphics[width=0.24\linewidth]{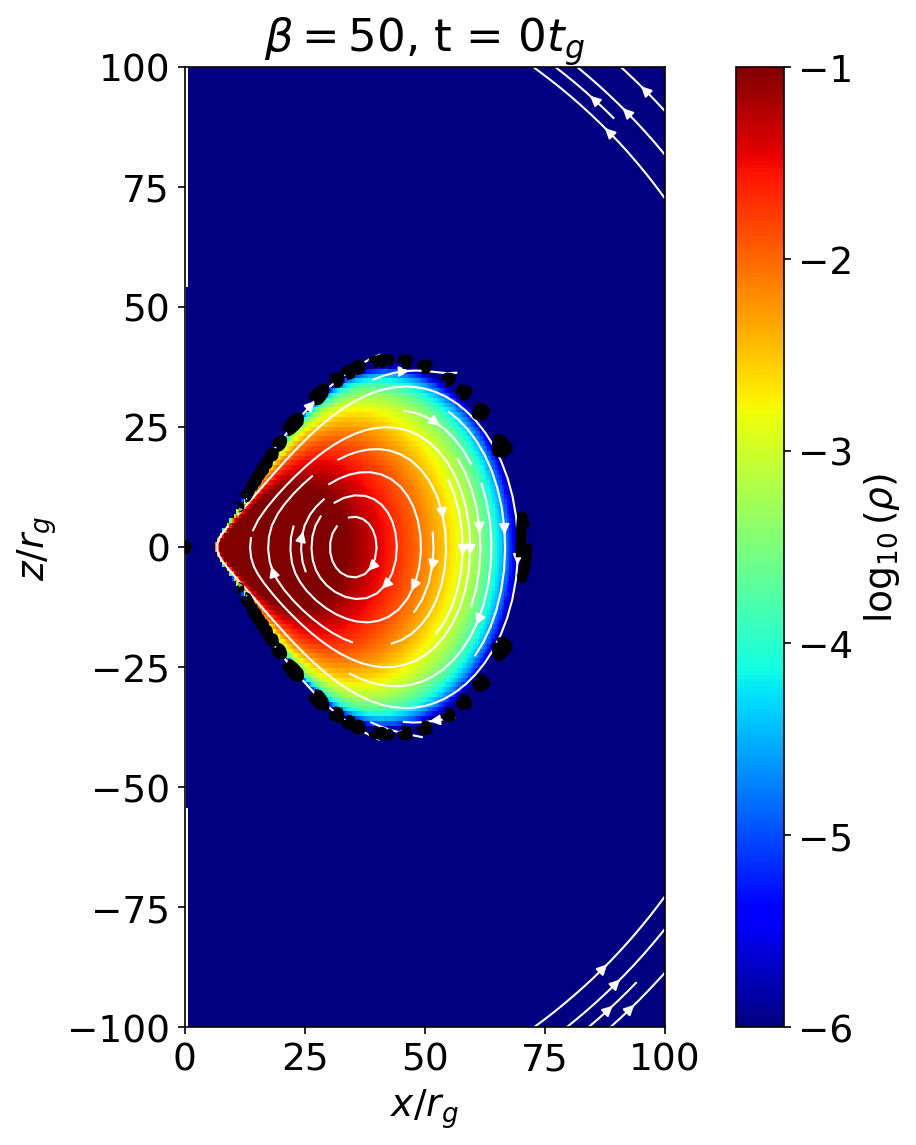}
    \includegraphics[width=0.24\linewidth]{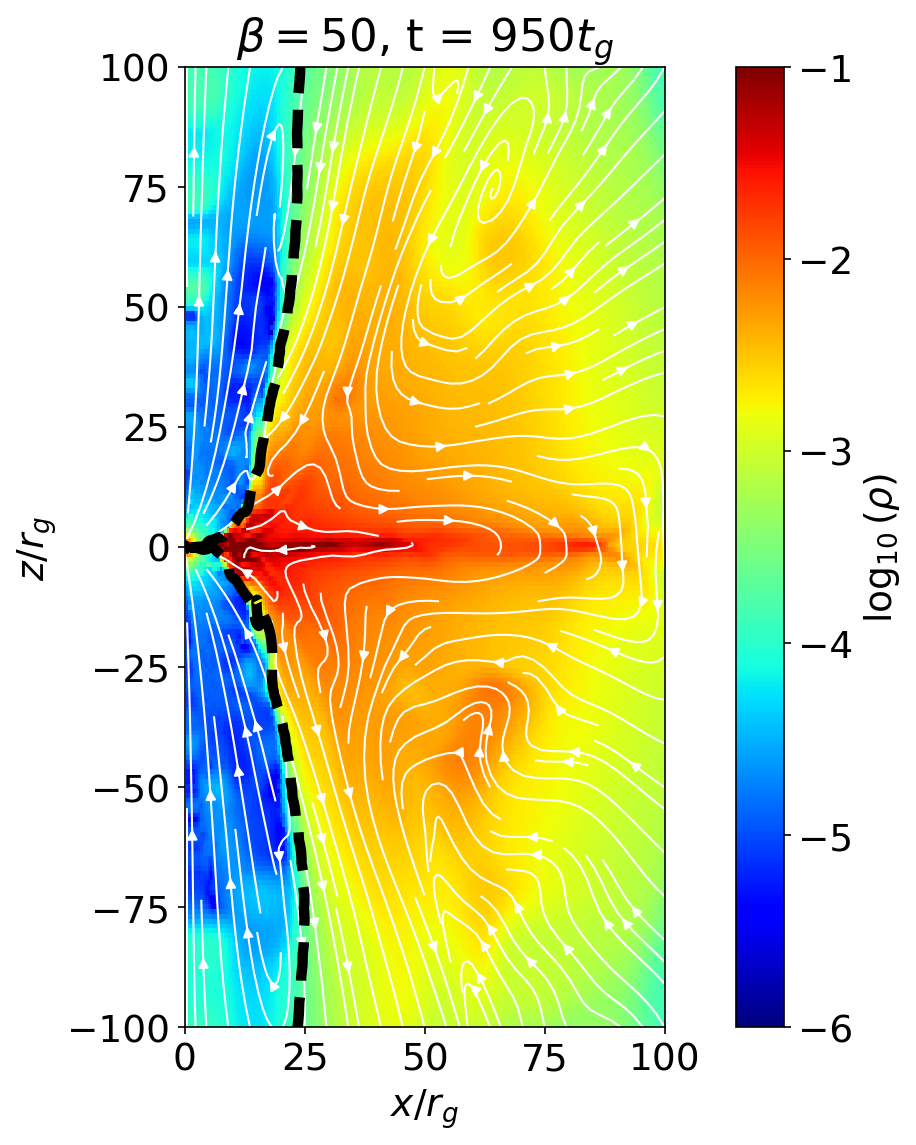}
    \includegraphics[width=0.24\linewidth]{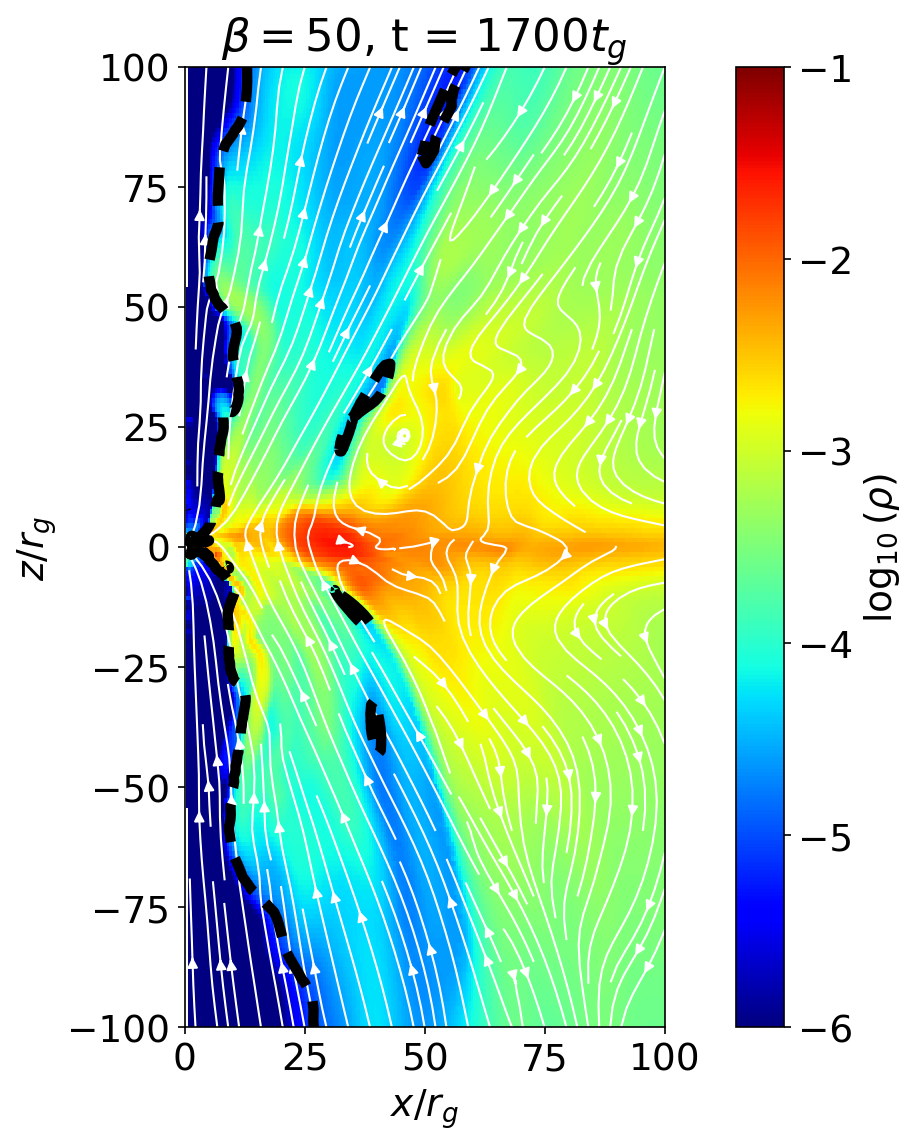}
    \includegraphics[width=0.24\linewidth]{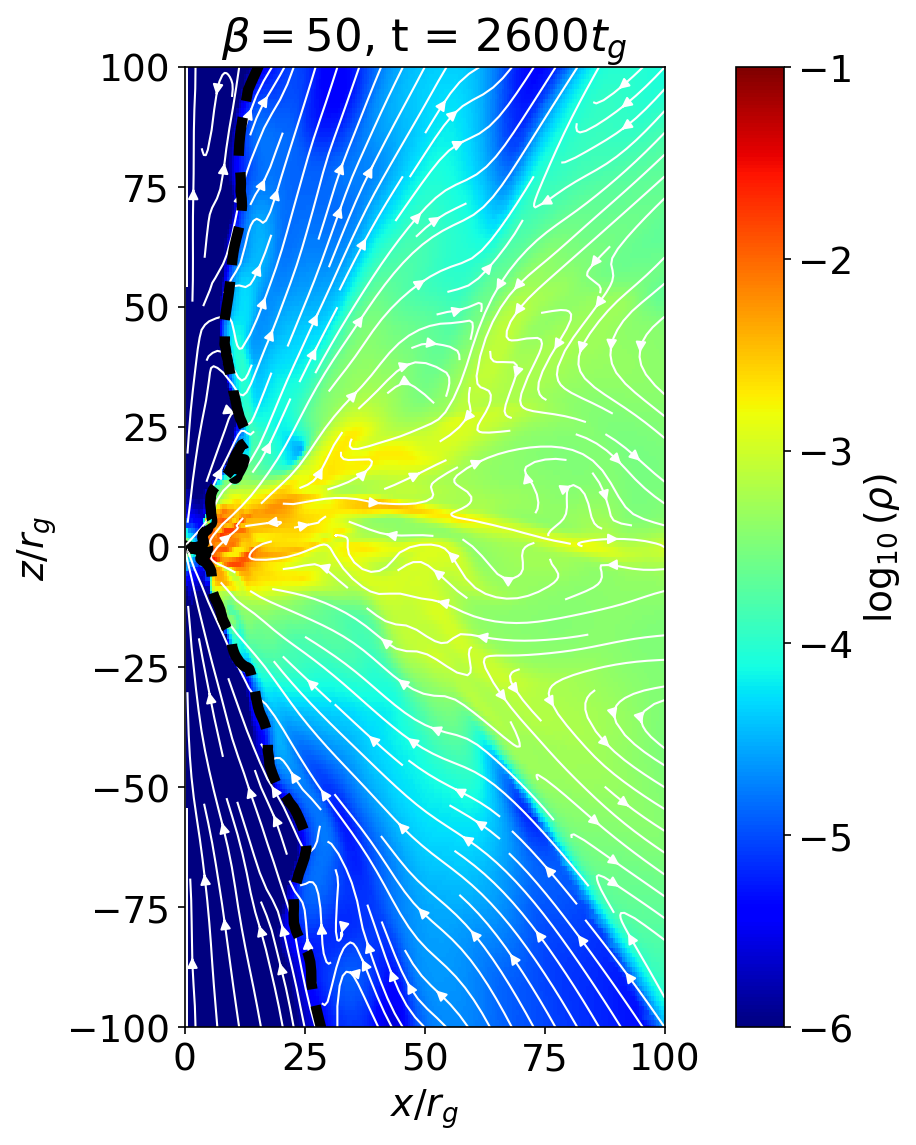}
    
    \includegraphics[width=0.24\linewidth]{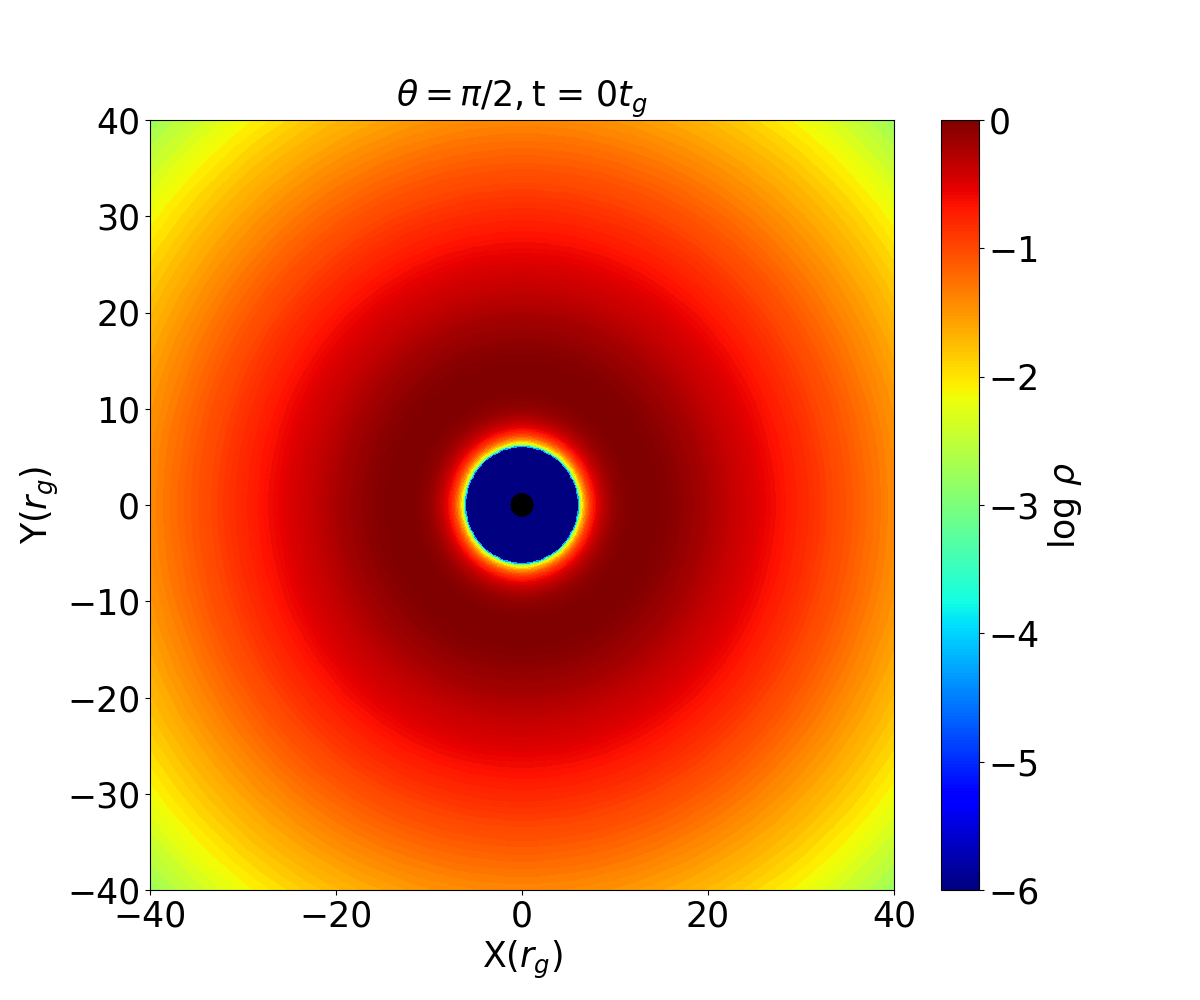}
    \includegraphics[width=0.24\linewidth]{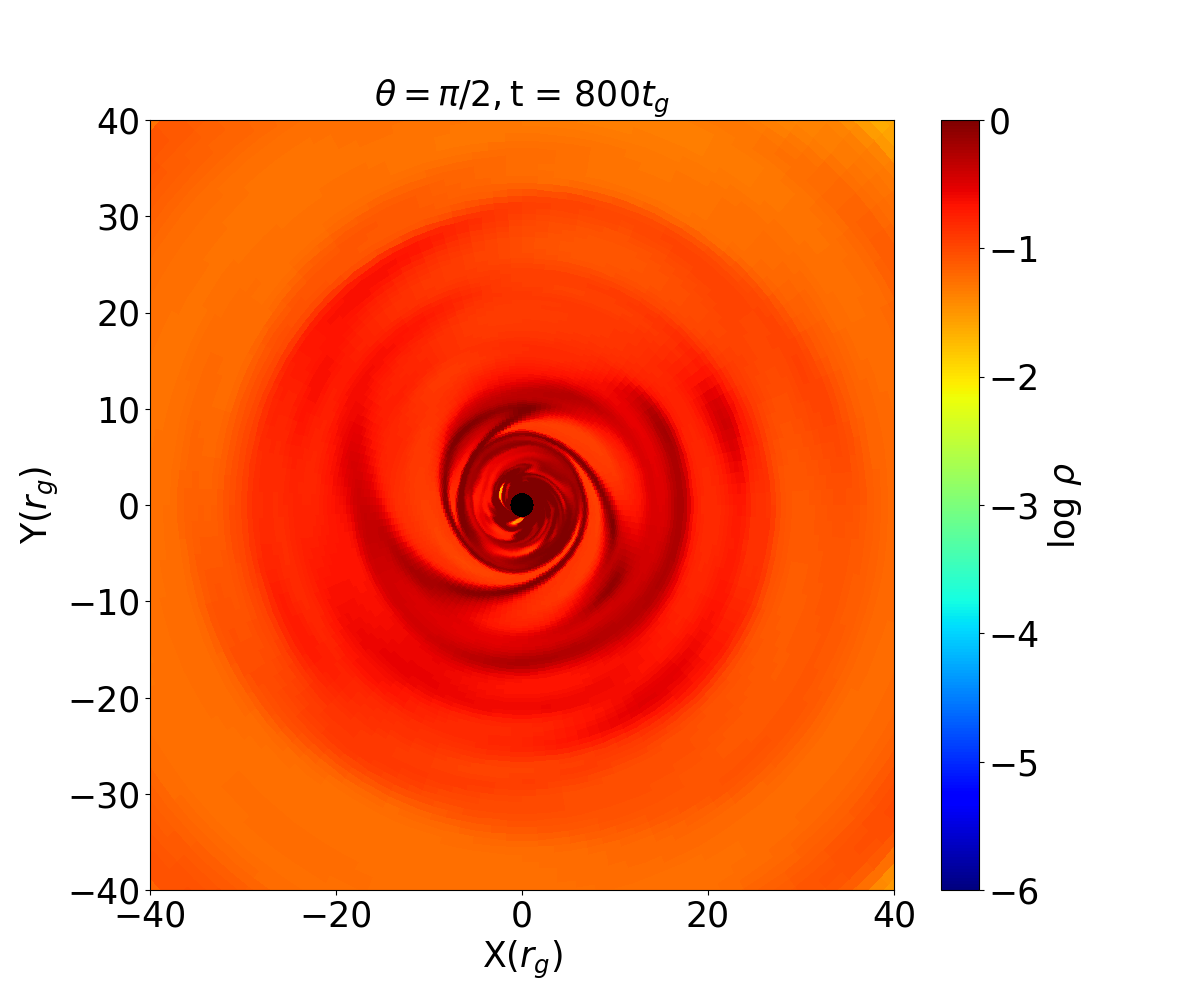}
    \includegraphics[width=0.24\linewidth]{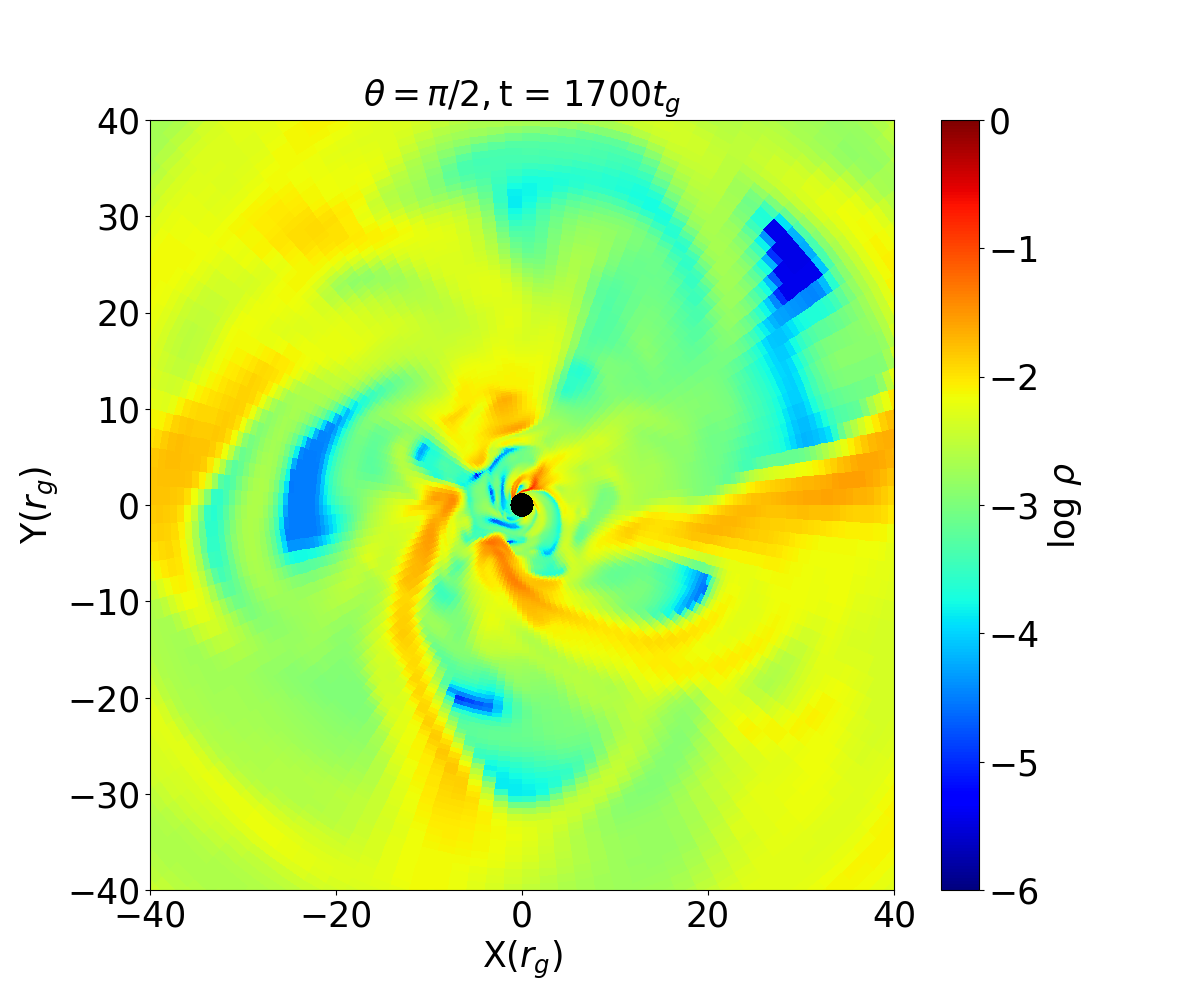}
    \includegraphics[width=0.24\linewidth]{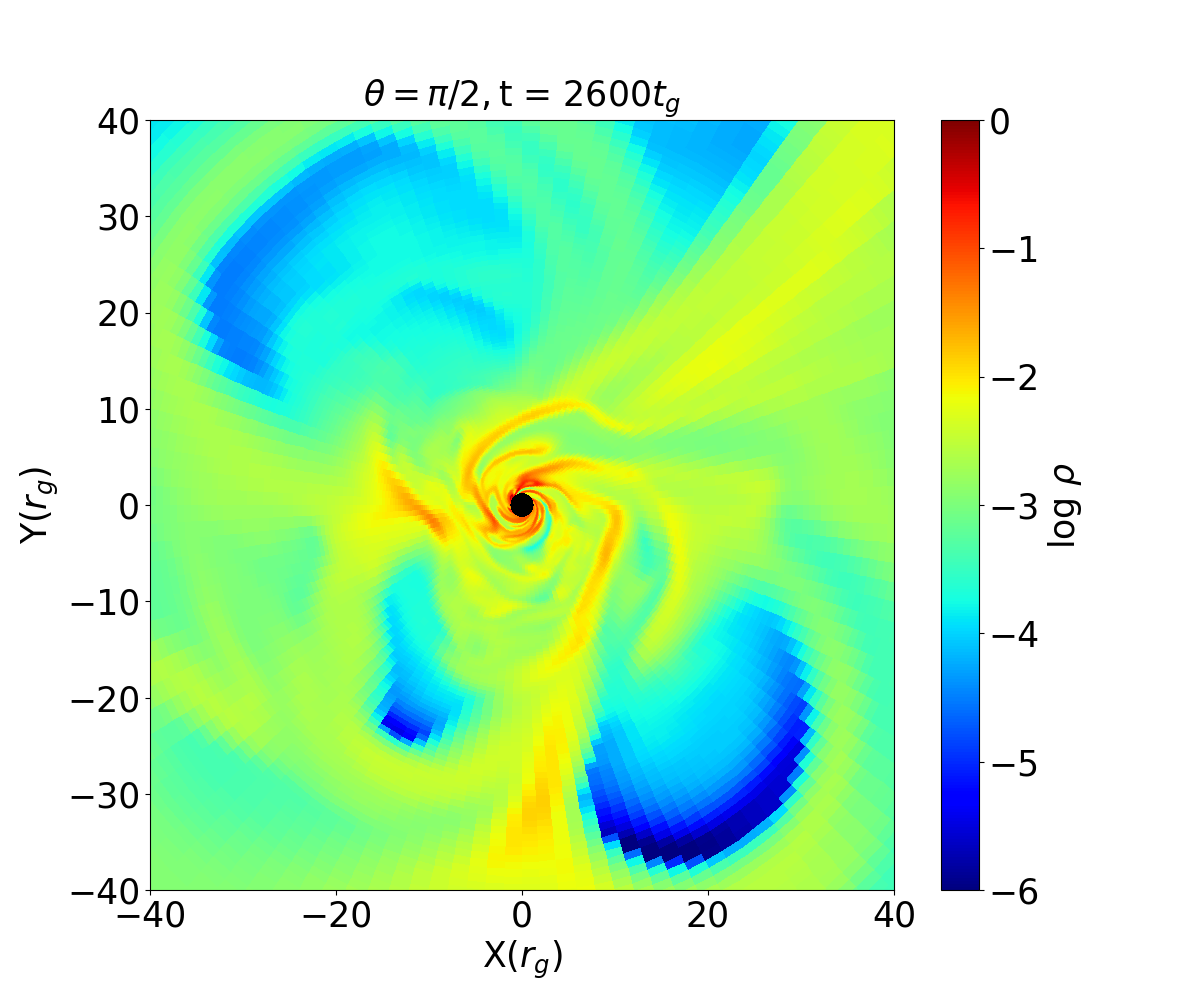}
    \caption{Evolution
 of torus density ($\rho$) for $\beta = 50$ with the vector potential $A_{\phi}^{(1)}$ and spin parameter $a = 0.9$. The top row shows density along the poloidal plane $(\phi = 0)$ overlaid with magnetic field streamlines (white lines), while the bottom row shows the equatorial plane $(\theta = \pi/2)$ at times $t = 0$, $800$, $1700$, and $2600\,t_g$.}
    \label{fig:beta50aphi1}
\end{figure}

Figure~\ref{fig:beta50aphi1} illustrates the evolution of the torus density for a plasma with $\beta = 50$, threaded by the poloidal magnetic field configuration $A_{\phi}^{(1)}$ around a black hole with spin $a = 0.9$. The top row shows density slices along the poloidal plane ($\phi = 0$) overlaid with magnetic field streamlines, highlighting the development and amplification of magnetic flux near the black hole. The black dashed contour in all 2D density maps represents the jet boundary, defined by the magnetization parameter $\sigma = 1$, which separates the magnetically dominated funnel region from the surrounding disk plasma.
 Initially, the field lines are weakly wound around the torus, but as the simulation progresses, the differential rotation shears the poloidal field, producing a stronger vertical field component near the inner disk and forming collimated field structures along the rotation axis.

\begin{figure}[H]
    \includegraphics[width=0.24\linewidth]{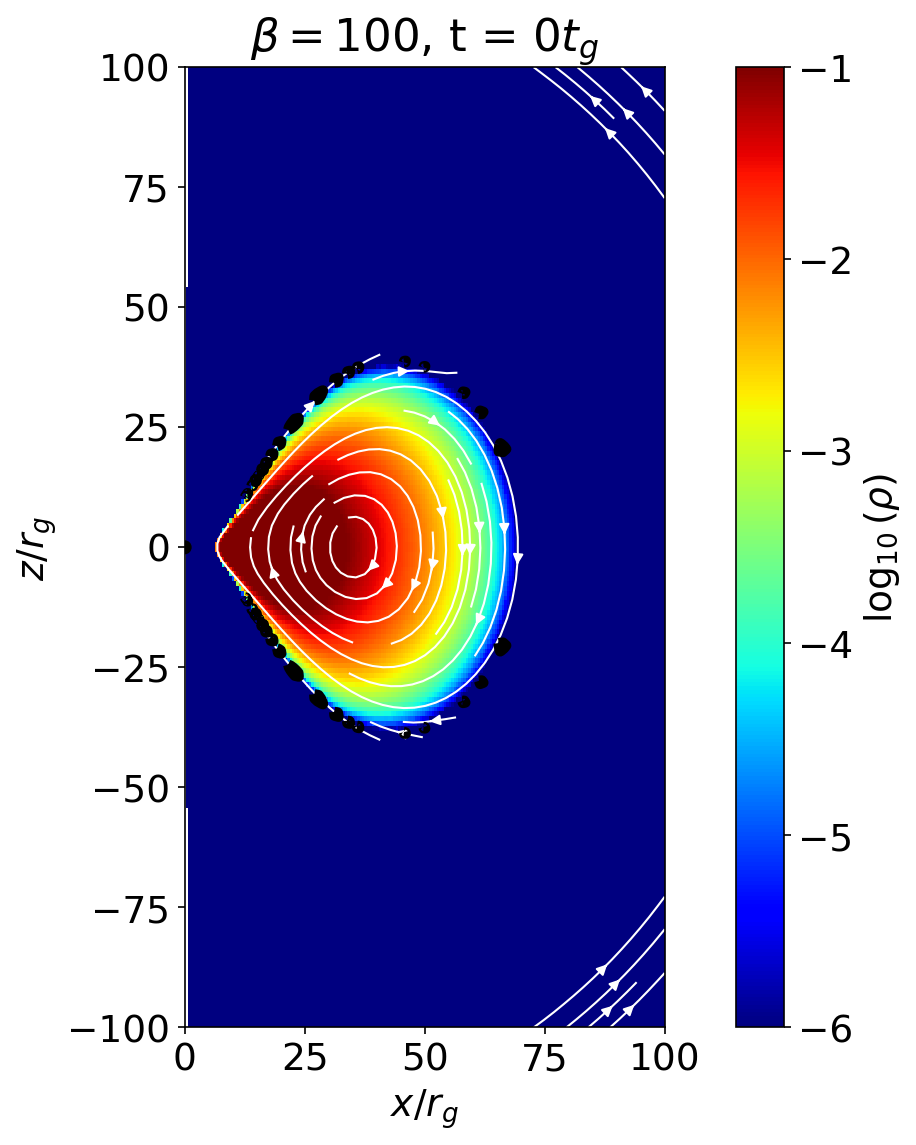}
    \includegraphics[width=0.24\linewidth]{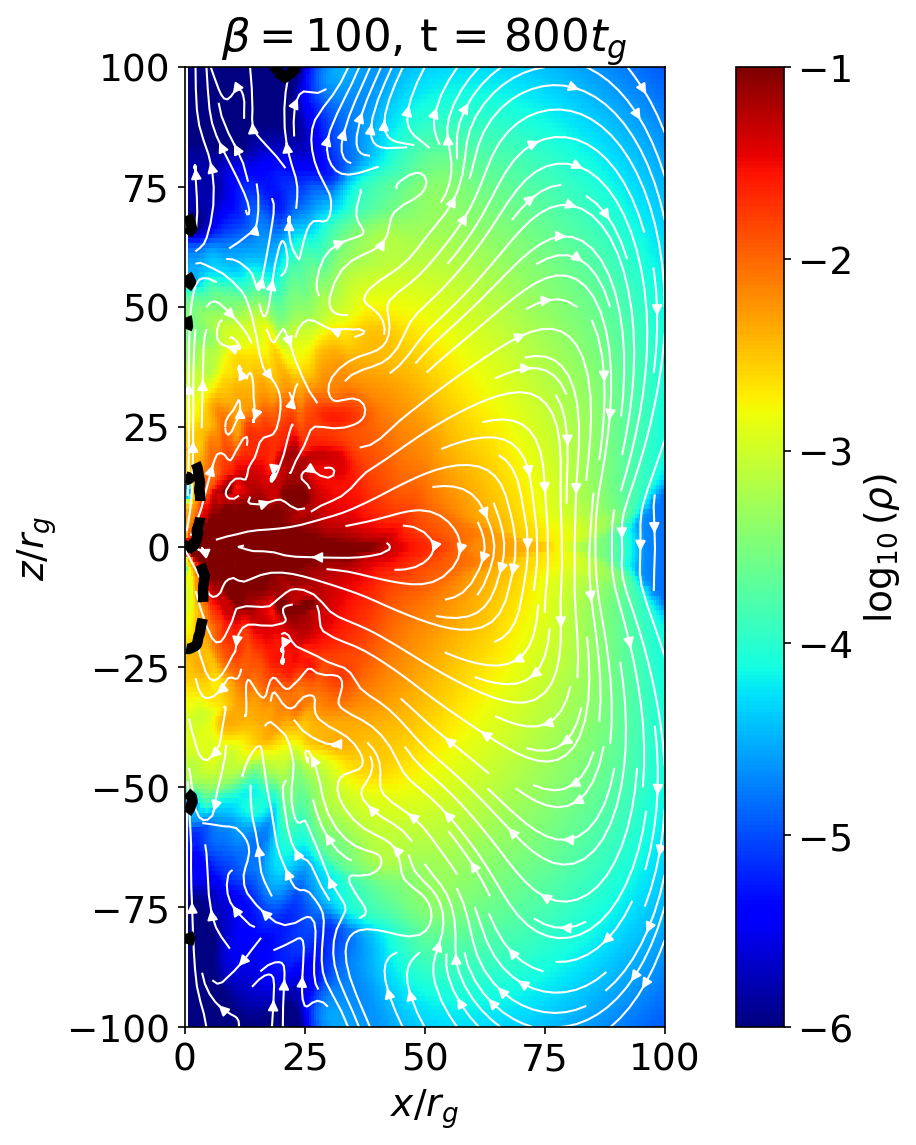}
    \includegraphics[width=0.24\linewidth]{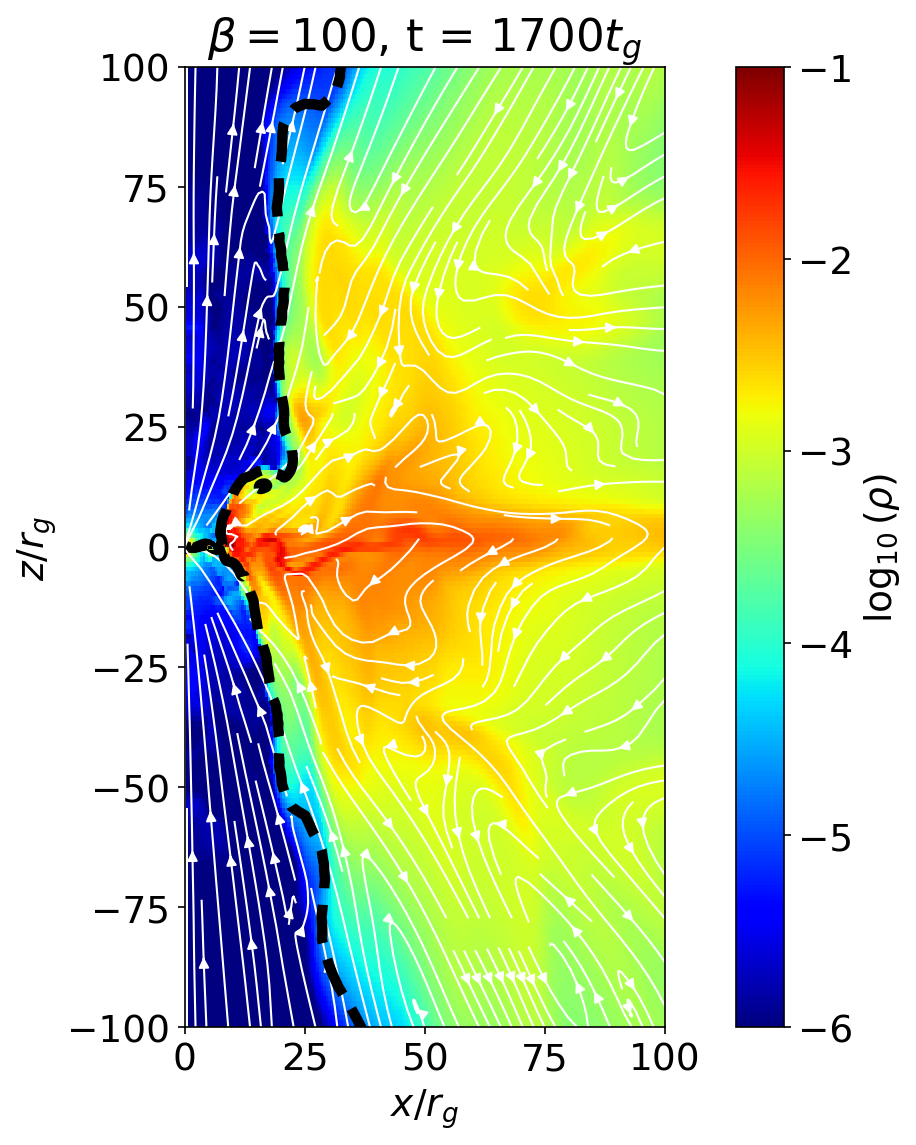}
    \includegraphics[width=0.24\linewidth]{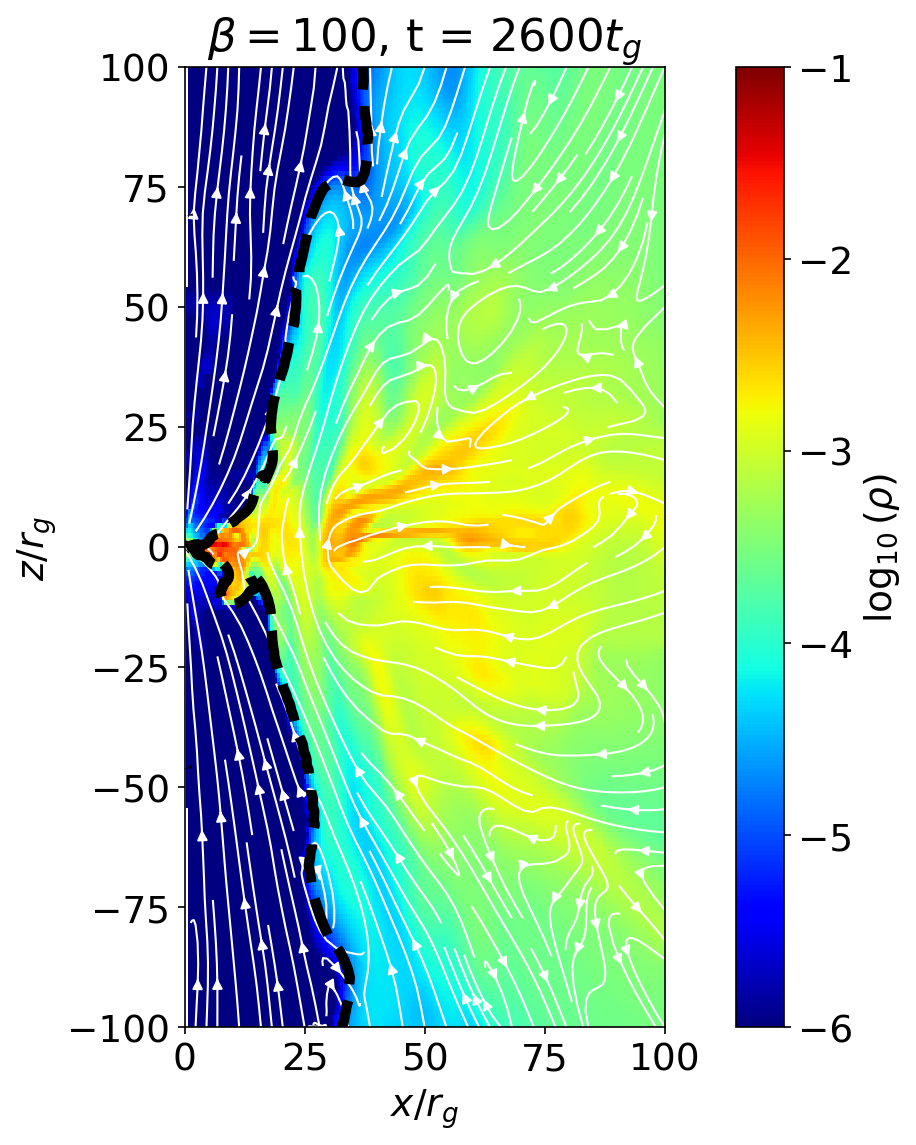}
    \caption{Evolution of torus density ($\rho$) for $\beta = 100$ with the vector potential $A_{\phi}^{(1)}$ and spin parameter $a = 0.9$. The plot shows density along the poloidal plane $(\phi = 0)$ overlaid with magnetic field streamlines at times $t = 0$, $800$, $1700$, and $2600\,t_g$.}
    \label{fig:beta100aphi1}
\end{figure}

The bottom panels show the corresponding slices in the equatorial plane ($\theta = \pi/2$), illustrating the orbital motion and gradual inward transport of matter. As the simulation evolves, the torus develops prominent spiral density structures driven by magnetorotational instability (MRI)-induced turbulence. The $\beta = 50$ model exhibits enhanced magnetic activity and stronger turbulent motions compared to the $\beta = 100$ case (Figure~\ref{fig:beta100aphi1}), where the flow remains relatively more laminar and the magnetic pressure is less dominant.

\begin{figure}[H]
    \includegraphics[width=0.24\linewidth]{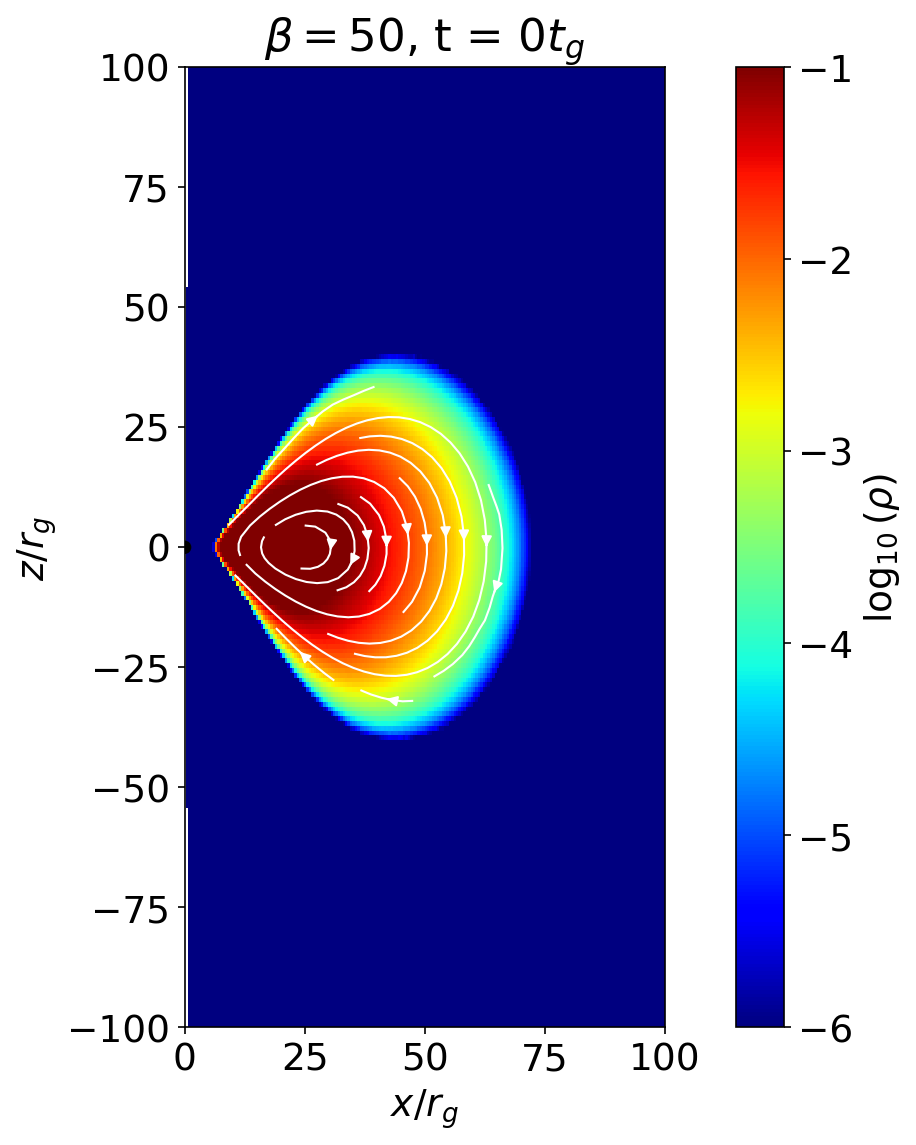}
    \includegraphics[width=0.24\linewidth]{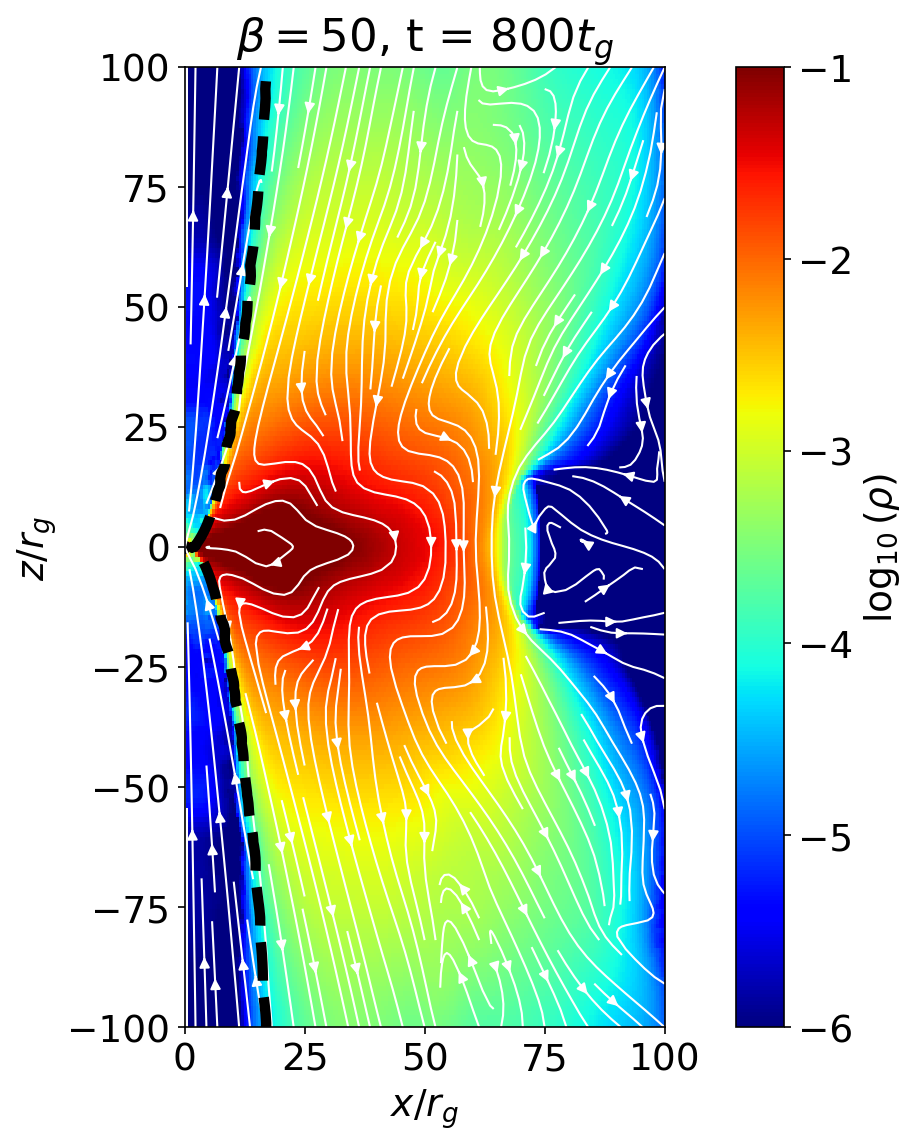}
    \includegraphics[width=0.24\linewidth]{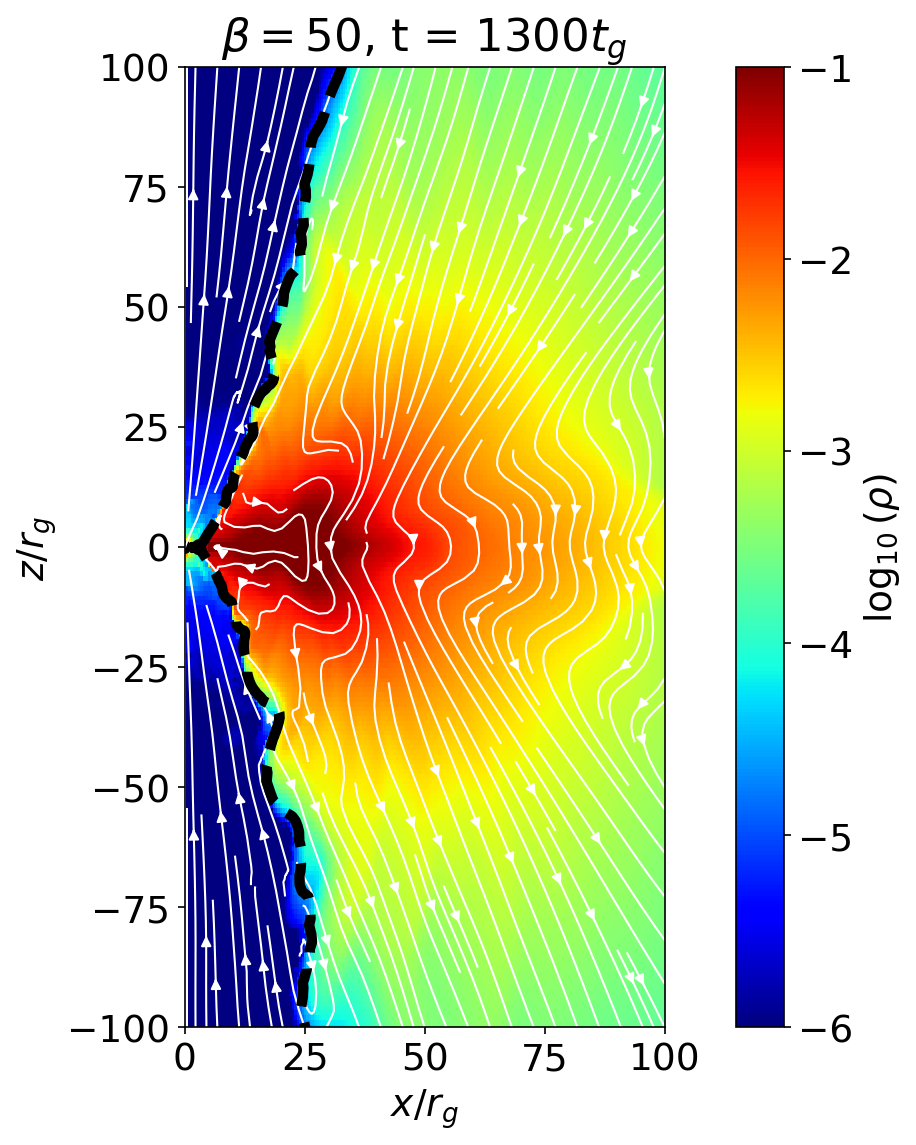}
    \includegraphics[width=0.24\linewidth]{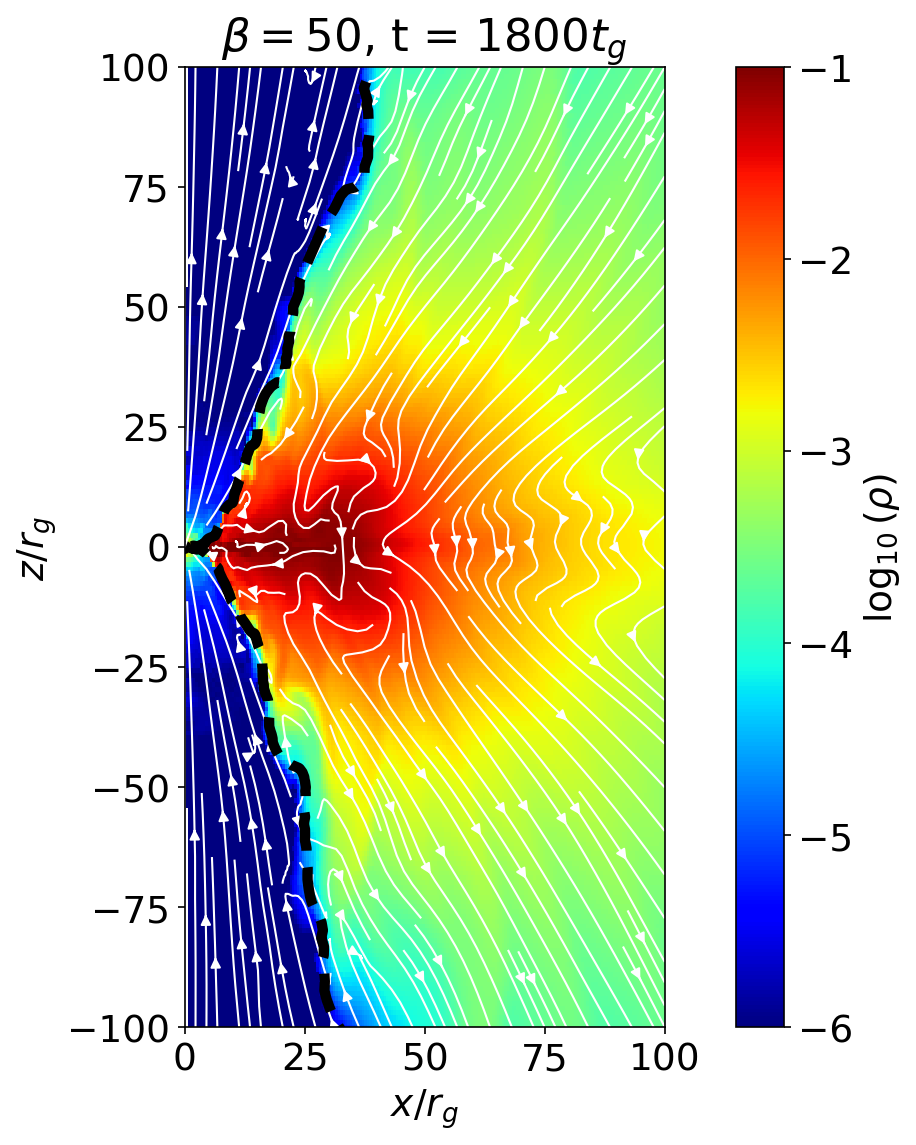}
     \caption{Evolution of torus density ($\rho$) for $\beta = 50$ with the vector potential $A_{\phi}^{(2)}$ and spin parameter $a = 0.935$. The plot shows density along the poloidal plane $(\phi = 0)$ overlaid with magnetic field streamlines at times $t = 0$, $800$, $1300$, and $1800\,t_g$.}
    \label{fig:beta50aphi2}
\end{figure}

\begin{figure}[H]
    \includegraphics[width=0.24\linewidth]{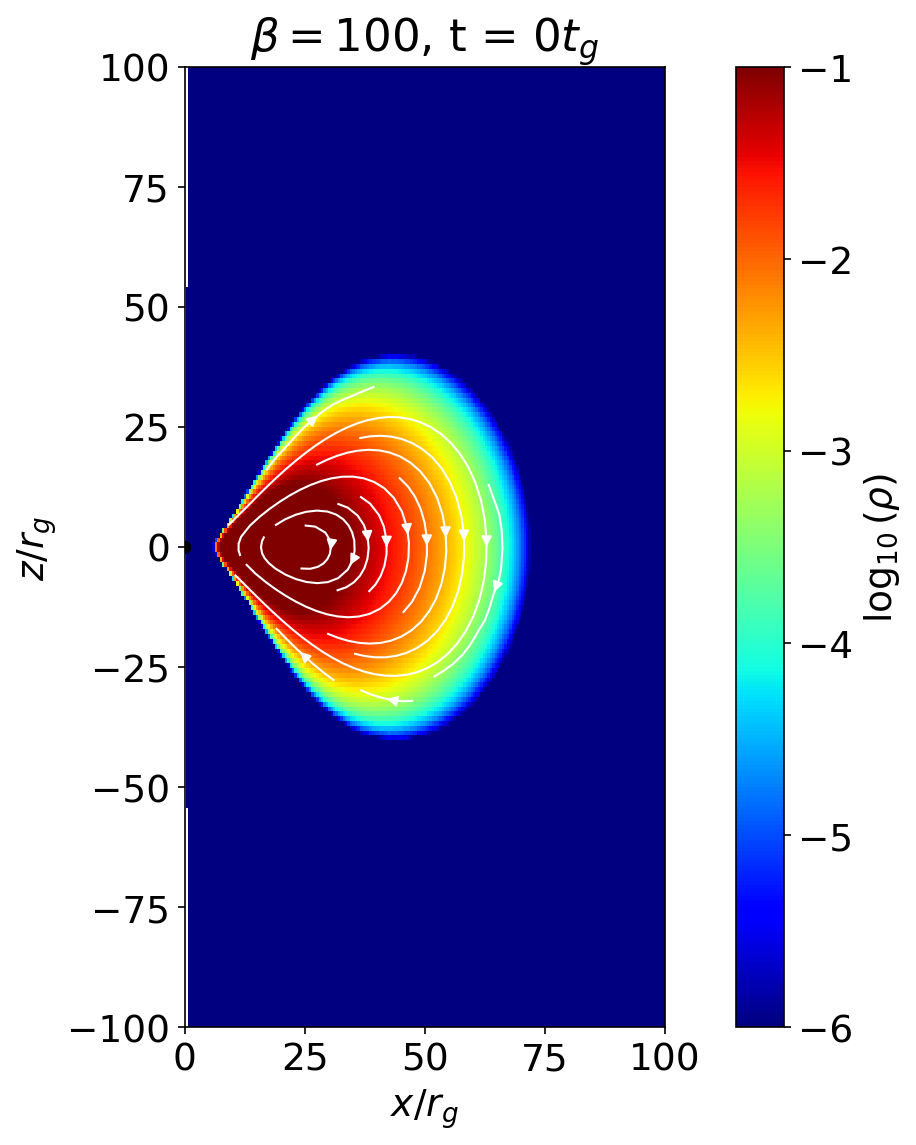}
    \includegraphics[width=0.24\linewidth]{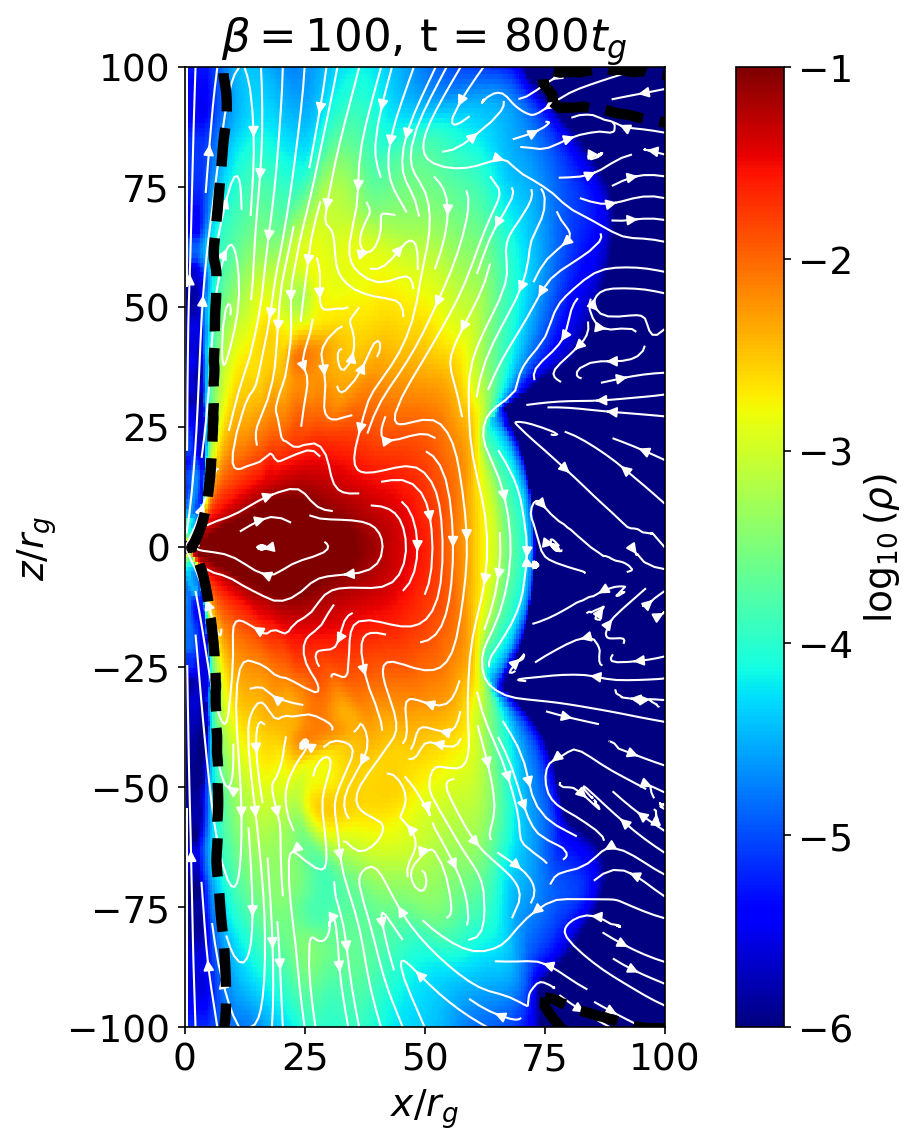}
    \includegraphics[width=0.24\linewidth]{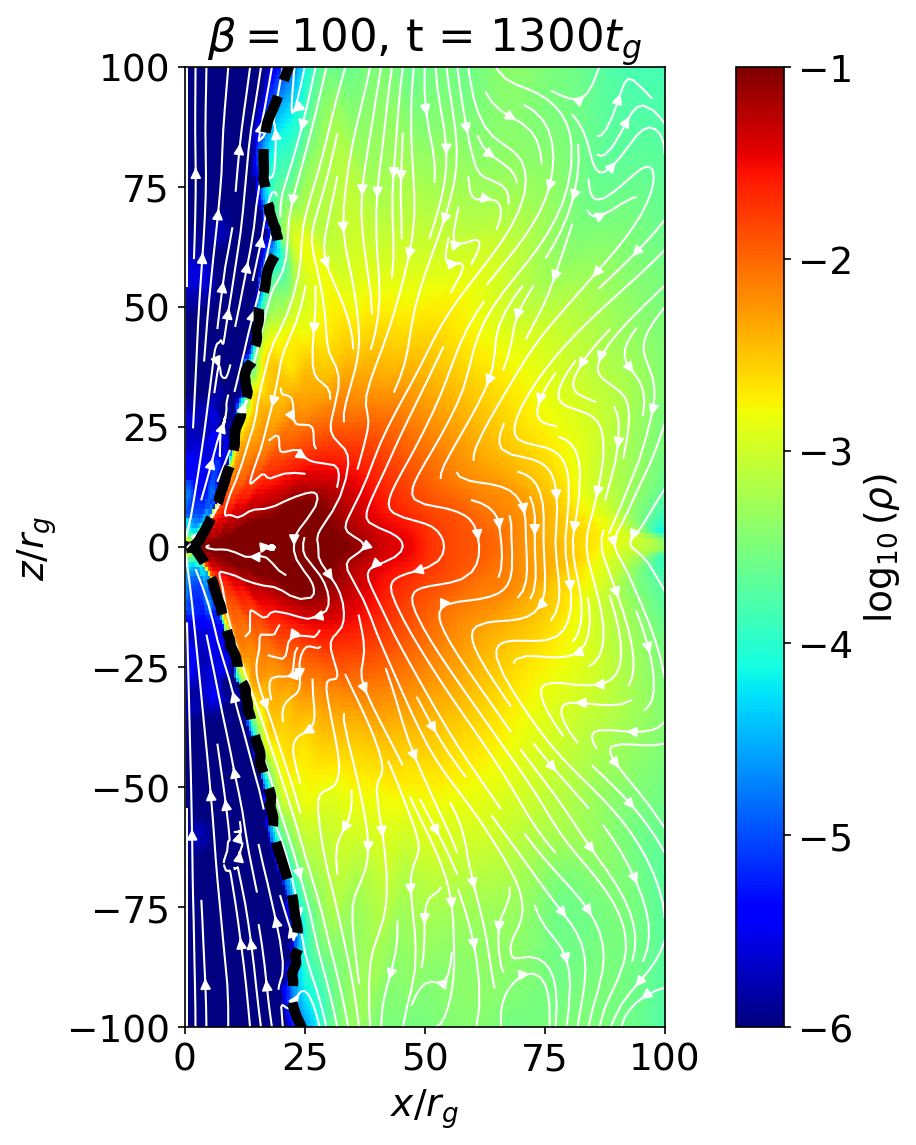}
    \includegraphics[width=0.24\linewidth]{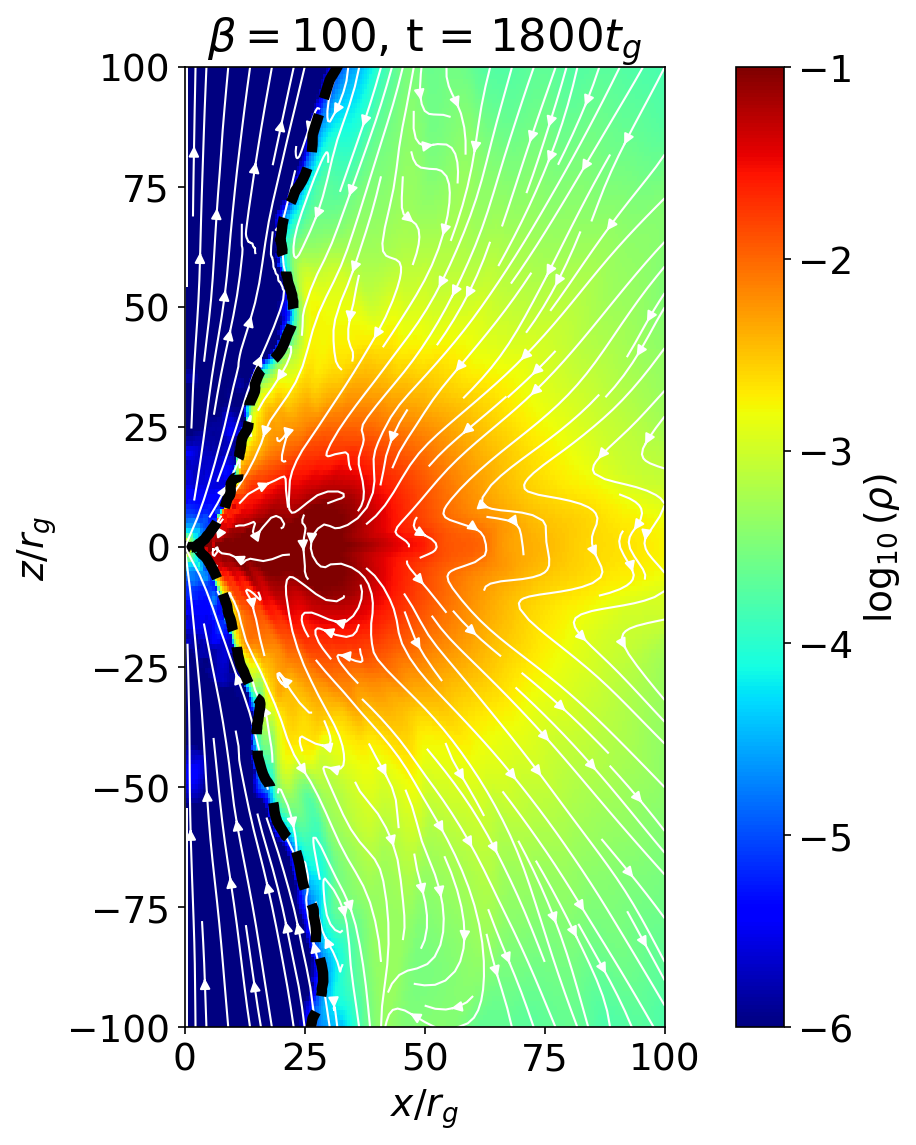}
    \caption{Evolution of torus density ($\rho$) for $\beta = 100$ with the vector potential $A_{\phi}^{(2)}$ and spin parameter $a = 0.935$. The plot shows density along the poloidal plane $(\phi = 0)$ overlaid with magnetic field streamlines at times $t = 0$, $800$, $1300$, and $1800\,t_g$.}
    \label{fig:beta100aphi2}
\end{figure}

Figures~\ref{fig:beta50aphi2} and~\ref{fig:beta100aphi2} present similar diagnostics for the second magnetic configuration, $A_{\phi}^{(2)}$, with $\beta = 50$ and $100$, respectively, and black hole spin $a = 0.935$. In these models, the field topology results in more efficient vertical collimation of magnetic field lines in the polar regions. $A_{\phi}^{(2)}$ supports a more symmetric and well-collimated outflow compared to $A_{\phi}^{(1)}$.
In particular, the jet cone is noticeably broader and more collimated for the $\beta = 50$ case, suggesting enhanced magnetic support and a stronger, magnetically dominated spine. Furthermore, the accretion torus in $A_{\phi}^{(1)}$ evolves more rapidly than in $A_{\phi}^{(2)}$, reaching jet launching phase at earlier times. This faster evolution can be attributed to the stronger initial poloidal flux and enhanced magnetic stresses in $A_{\phi}^{(1)}$, which promote more efficient angular momentum transport and rapid advection of magnetic flux toward the black hole.

\vspace{-10pt}

\begin{figure}[H]
   \centering \textbf{$\beta=50$} \\[1mm] 
    \begin{minipage}{0.34\linewidth}
        \includegraphics[width=\linewidth]{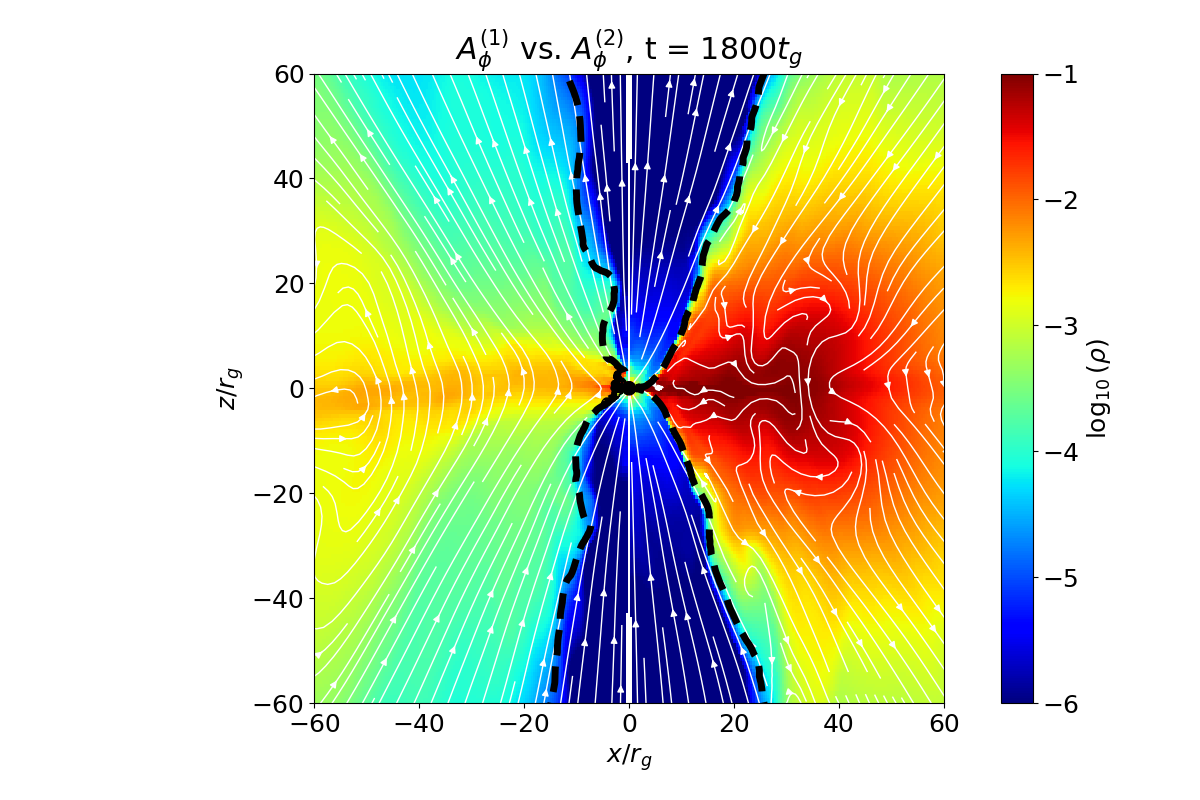}
    \end{minipage}%
    \begin{minipage}{0.34\linewidth}
        \includegraphics[width=\linewidth]{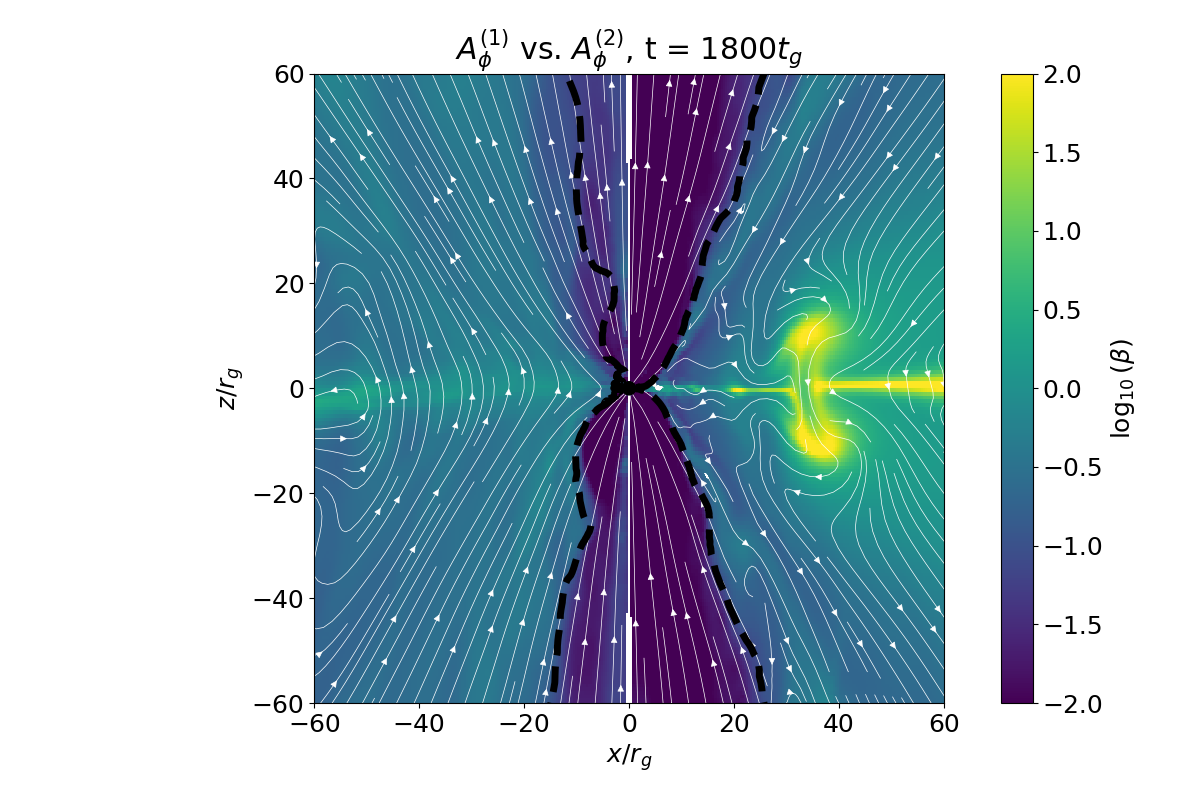}
    \end{minipage}%
    \begin{minipage}{0.34\linewidth}
        \includegraphics[width=\linewidth]{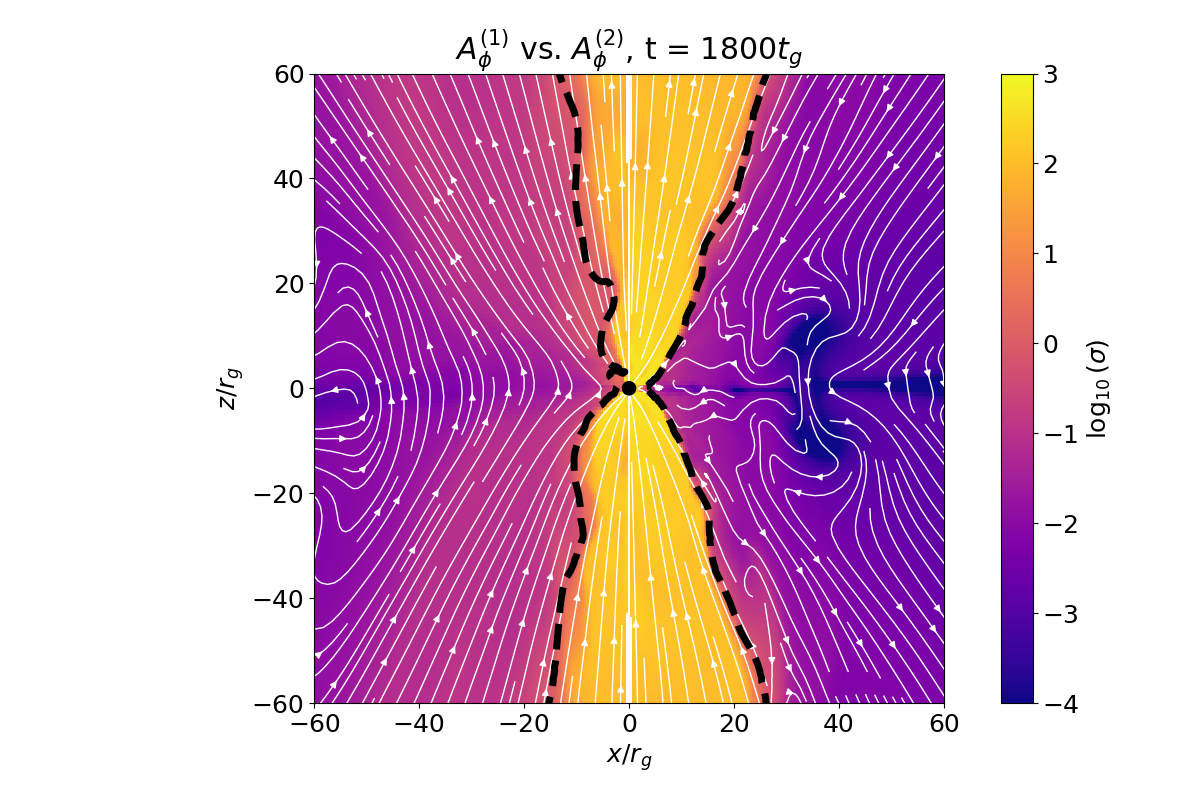}
    \end{minipage}

    \vspace{1mm} 

    \textbf{$\beta=100$} \\[1mm] 
    \begin{minipage}{0.34\linewidth}
        \includegraphics[width=\linewidth]{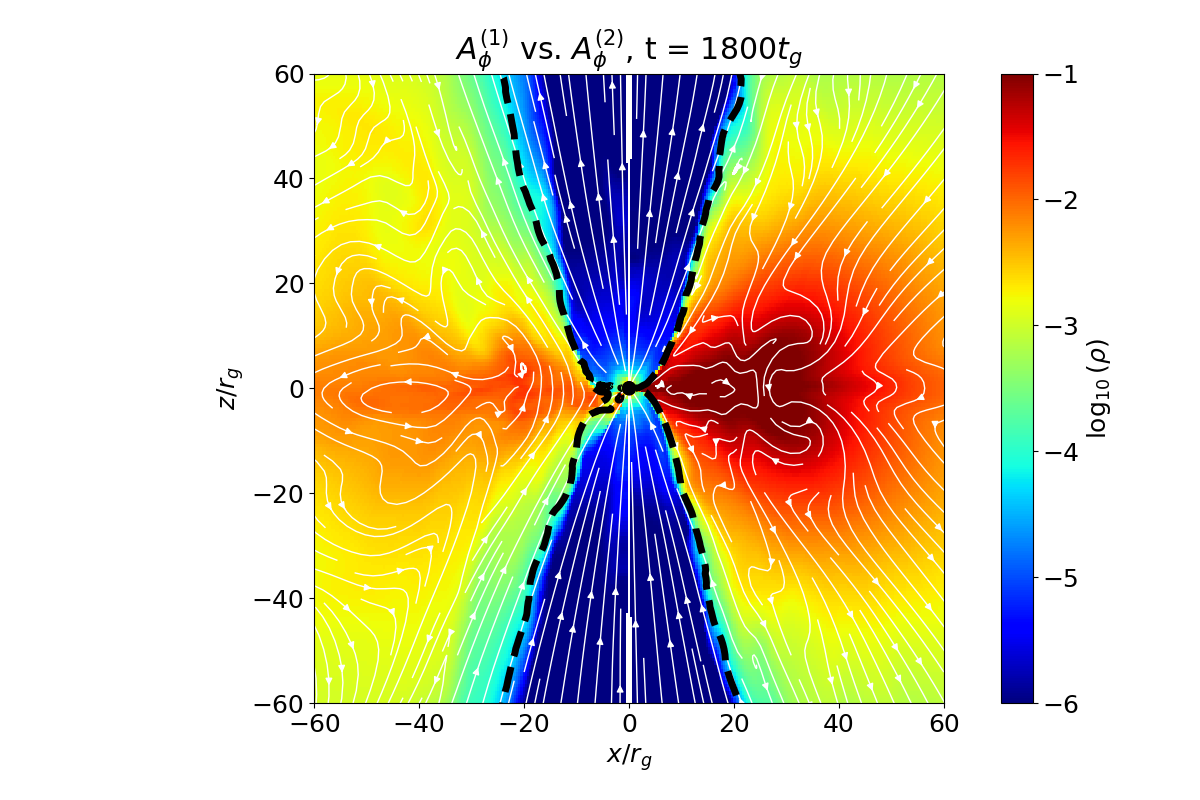}
    \end{minipage}%
    \begin{minipage}{0.34\linewidth}
        \includegraphics[width=\linewidth]{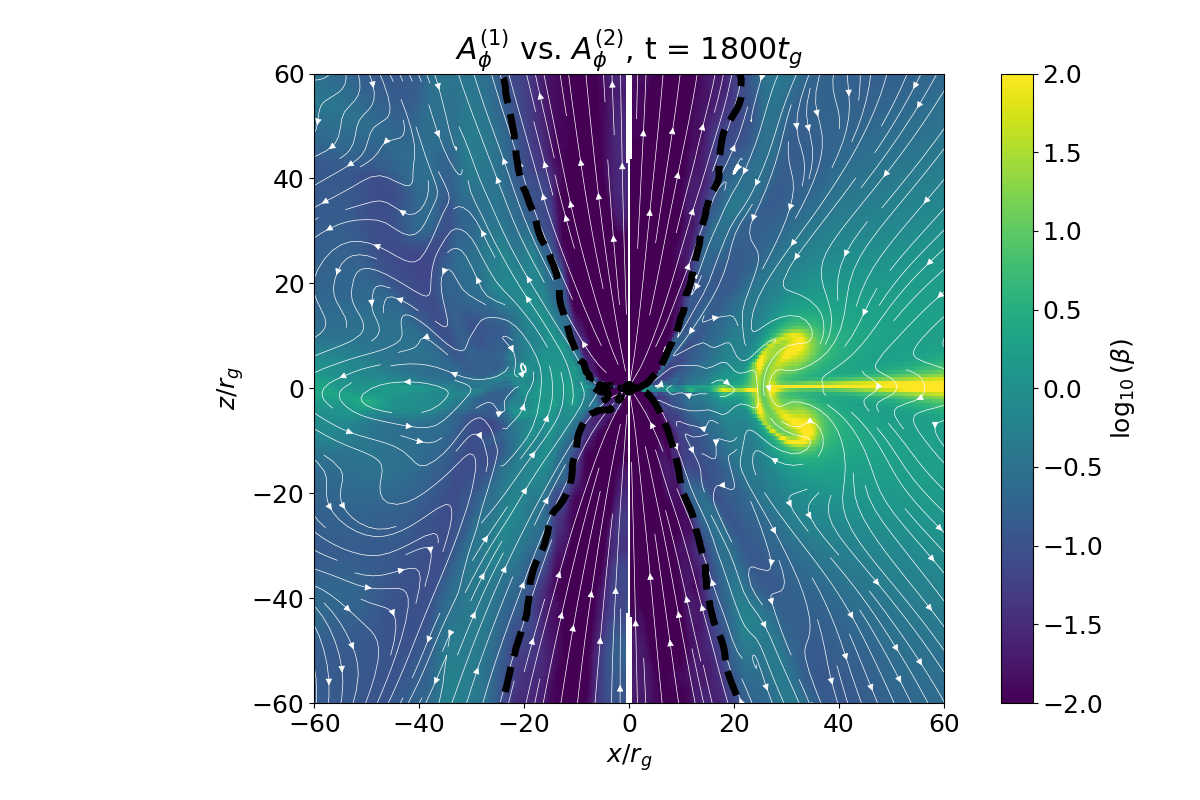}
    \end{minipage}%
    \begin{minipage}{0.34\linewidth}
        \includegraphics[width=\linewidth]{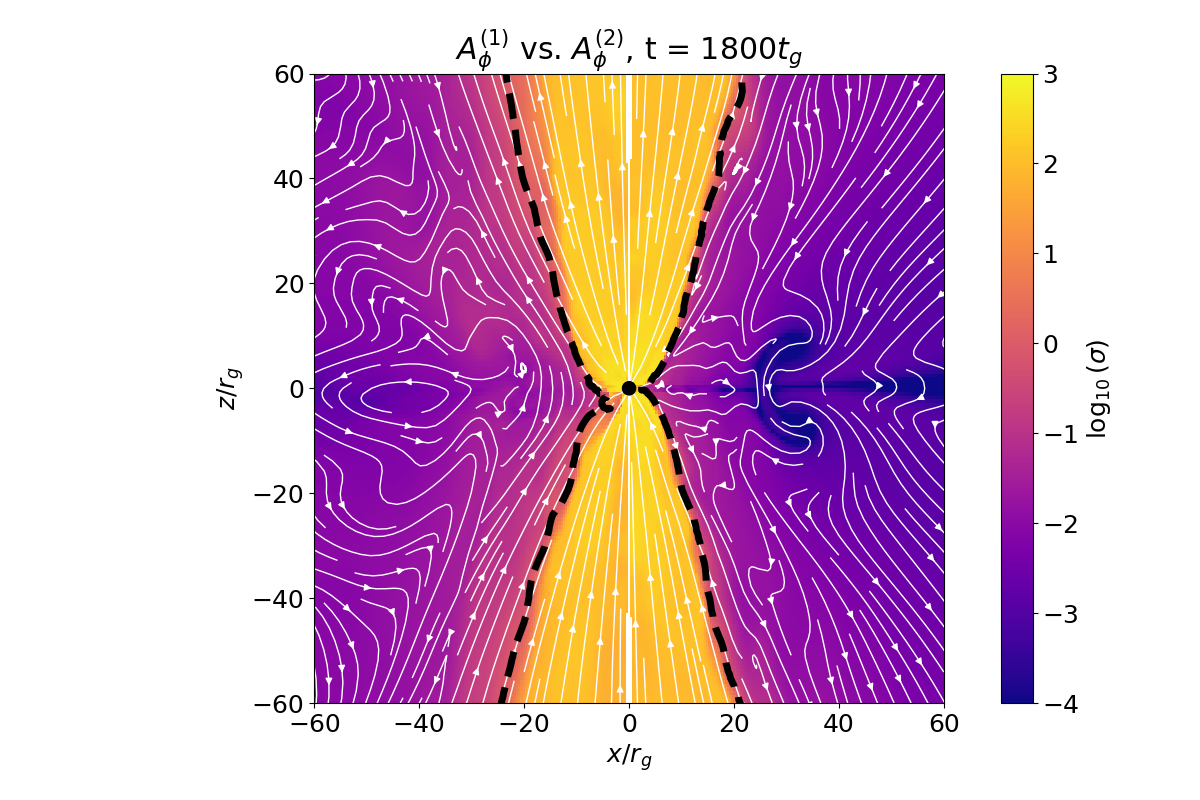}
    \end{minipage}

    \caption{Snapshots 
 at $t = 1800$ showing the spatial distributions of key quantities for the two magnetic configurations. The left and right columns correspond to $A_{\phi}^{(1)}$ and $A_{\phi}^{(2)}$, respectively in each quantity. The three panels in each row show (from left to right) the density, plasma $\beta$, and magnetization parameter $\sigma$. The \textbf{top row} corresponds to an initial plasma $\beta = 50$, and \mbox{the \textbf{bottom row}} to $\beta = 100$. White streamlines trace the poloidal magnetic field lines, illustrating the magnetic topology and its evolution in each configuration.}
    \label{aphi_1vs2}
\end{figure}

Figure~\ref{aphi_1vs2} illustrates the differences in the torus density, magnetic field ordering, magnetization, and jet/disk structure between the two configurations at the same evolutionary time. In the density panels of Figure~\ref{aphi_1vs2}, it is evident that the accretion torus evolves differently for the two magnetic configurations. For $A_{\phi}^{(1)}$, the torus is almost accreted by $t = 1300$, whereas in $A_{\phi}^{(2)}$ a significant fraction of the disk remains, indicating slower but more stable accretion. The effect is more visible for higher magnetization (lower plasma $\beta$); specifically, for $\beta = 50$, the $A_{\phi}^{(1)}$ torus accretes more rapidly compared to $\beta = 100$. The jet morphology also reflects these differences: for both configurations, the jet cone is broader and less collimated at $\beta = 50$ than at $\beta = 100$, as the stronger magnetic flux drives more powerful yet less collimated outflows, since the enhanced magnetic pressure near the black hole inflates the jet funnel and widens the outflow boundary.
Comparing the two initial vector potentials (Equations~(\ref{Aphi1}) and (\ref{Aphi2})), $A_{\phi}^{(1)}$ with its simple power-law radial dependence produces faster but more turbulent accretion and less ordered jet boundaries, while $A_{\phi}^{(2)}$, with its radially tapered, $\sin^3\theta$-weighted structure, leads to a more coherent magnetic field, smoother jet spine, and broader polar funnel. 

In addition to the density panels, the plasma $\beta$ and magnetization $\sigma$ plots in Figure~\ref{aphi_1vs2} further illustrate the differences in magnetic-field structure between the two configurations. The $A_{\phi}^{(2)}$ runs exhibit a more ordered and coherent magnetic field throughout the disk and polar regions, reflecting the large-scale poloidal geometry imposed by the initial vector potential. In contrast, $A_{\phi}^{(1)}$, which begins with stronger magnetization and larger initial flux, rapidly advects the magnetic field toward the black hole and therefore reaches the strong-flux (MAD) regime earlier. The field becomes so sufficiently strong in the $A_{\phi}^{(1)}$ case that magnetic pressure expands laterally, reducing jet collimation.

In all of our simulations, both $A_{\phi}^{(1)}$ and $A_{\phi}^{(2)}$ ultimately reach a MAD state; the key difference lies in the timescale and manner of flux accumulation. The $A_{\phi}^{(1)}$ configuration achieves MAD promptly due to rapid flux advection, whereas the $A_{\phi}^{(2)}$ configuration develops MAD more gradually, following a slower, disk-mediated buildup of \mbox{magnetic flux}.

These behaviors highlight how the initial vector-potential geometry can shape GRMHD simulations toward either rapid jet-launching studies or longer-term investigations of disk evolution and MAD onset.

\begin{figure}[H]
    \includegraphics[width=0.24\linewidth]{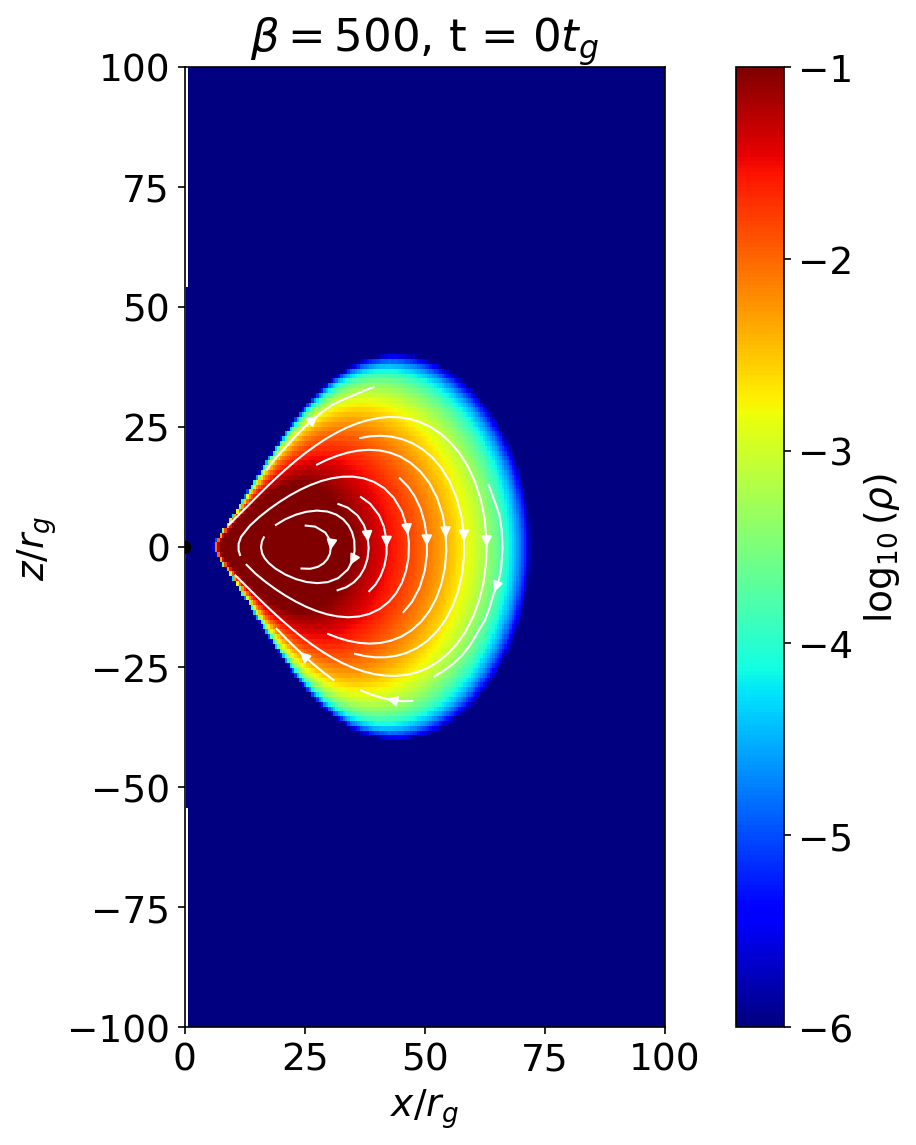}
    \includegraphics[width=0.24\linewidth]{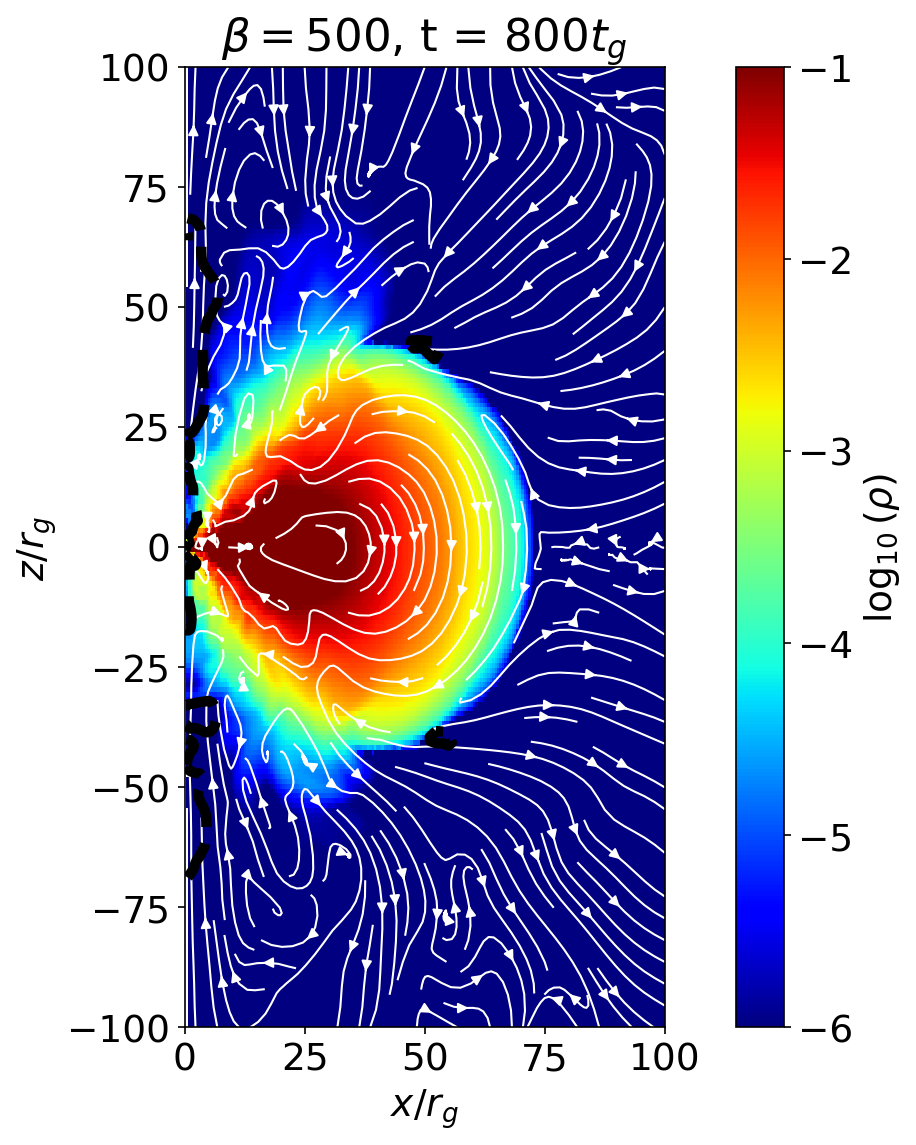}
    \includegraphics[width=0.24\linewidth]{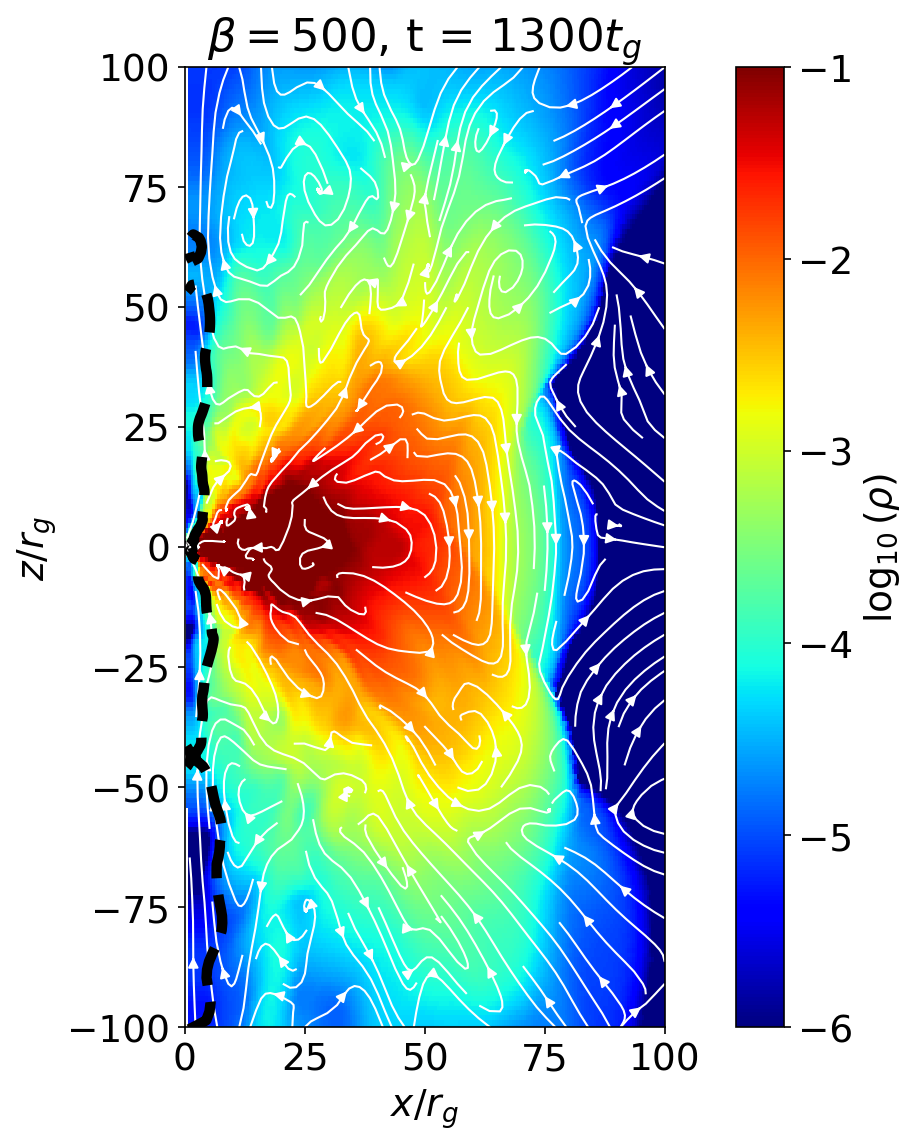}
    \includegraphics[width=0.24\linewidth]{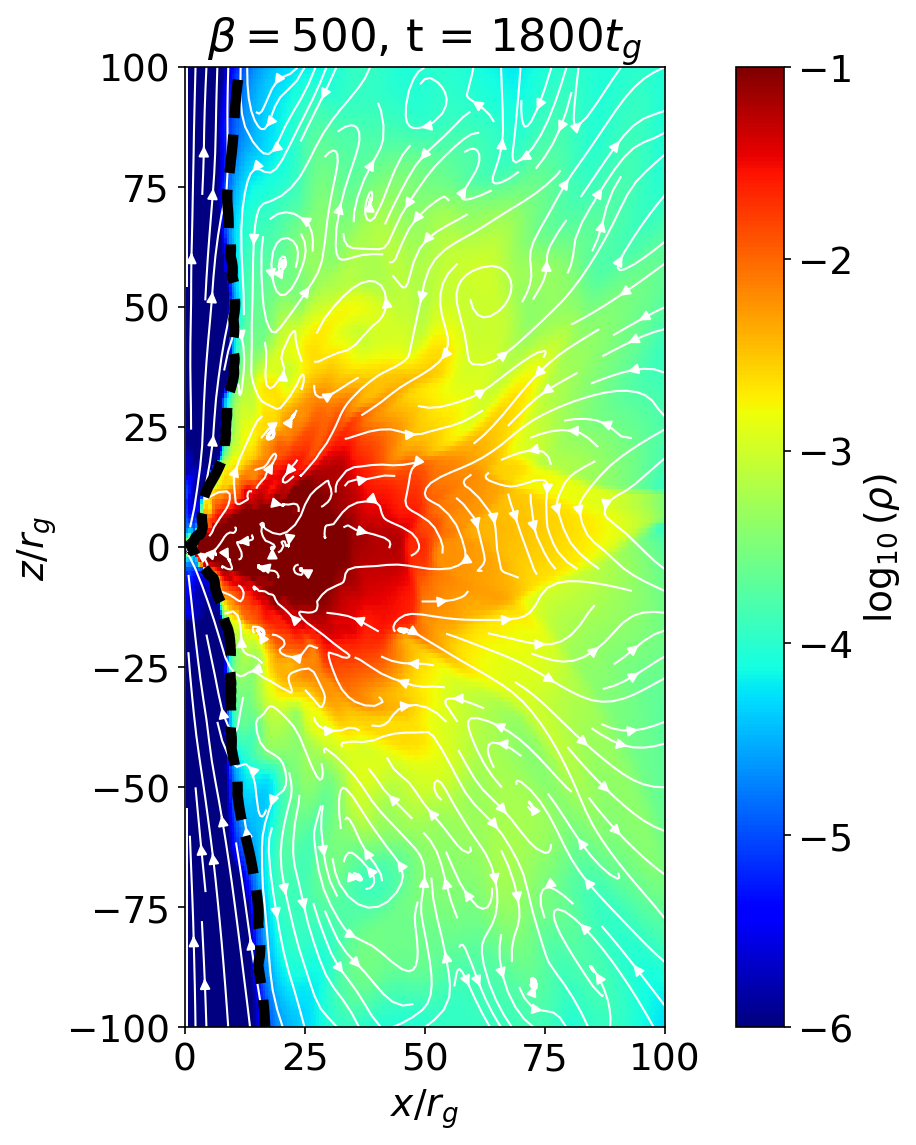}
    \caption{Torus density ($\rho$) evolution along a poloidal plane $(\phi = 0)$ with magnetic field streamlines at $t = 0$, $800$, $1300$, and $1800\,t_g$. for $\beta = 500$, due to the vector potential $A_{\phi}^{(2)}$ and spin parameter $a = 0.935$.}
    \label{fig:beta500aphi2}
\end{figure}

Figure ~\ref{fig:beta500aphi2} shows the evolution of the accretion torus for the weakly magnetized case with $\beta = 500$ using the vector potential configuration $A_{\phi}^{(2)}$ and black hole spin $a = 0.935$. In this regime, the magnetic field is dynamically subdominant, resulting in a flow that remains largely hydrodynamic throughout the simulation. The poloidal slices show distortion of the magnetic field lines and no evidence of strong vertical collimation or jet formation. The absence of a well-defined low-density funnel or significant magnetic flux accumulation near the black hole indicates that the system fails to reach a MAD state. Instead, the flow remains in a SANE-like configuration, dominated by gas pressure and exhibiting minimal magnetic feedback on the accretion dynamics.

\subsection{Time Evolution of Accretion and Magnetic Flux}
\label{3.3}

\begin{figure}[H]
    \includegraphics[width=0.48\linewidth]{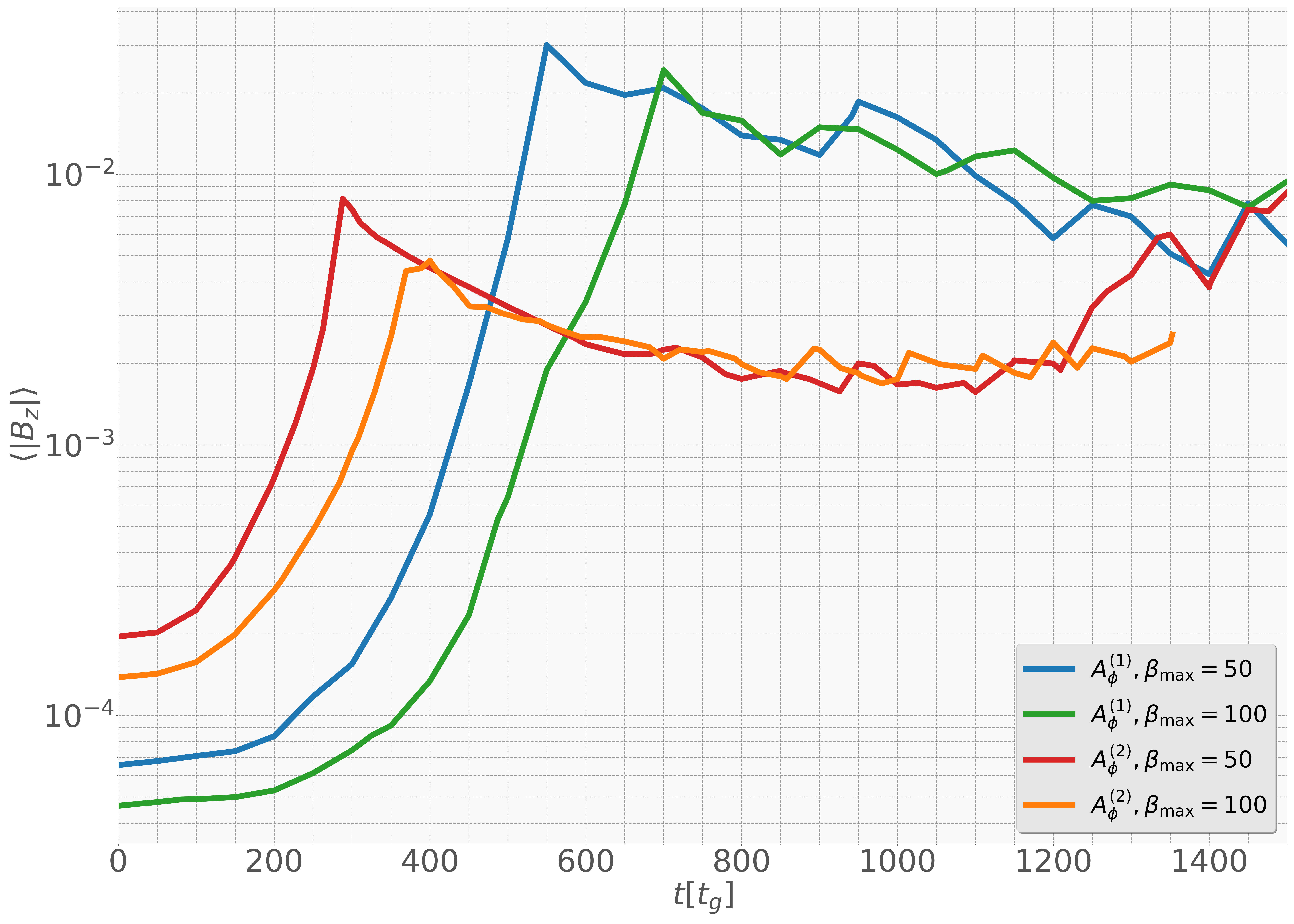}
    \hfill
     \includegraphics[width=0.48\linewidth]{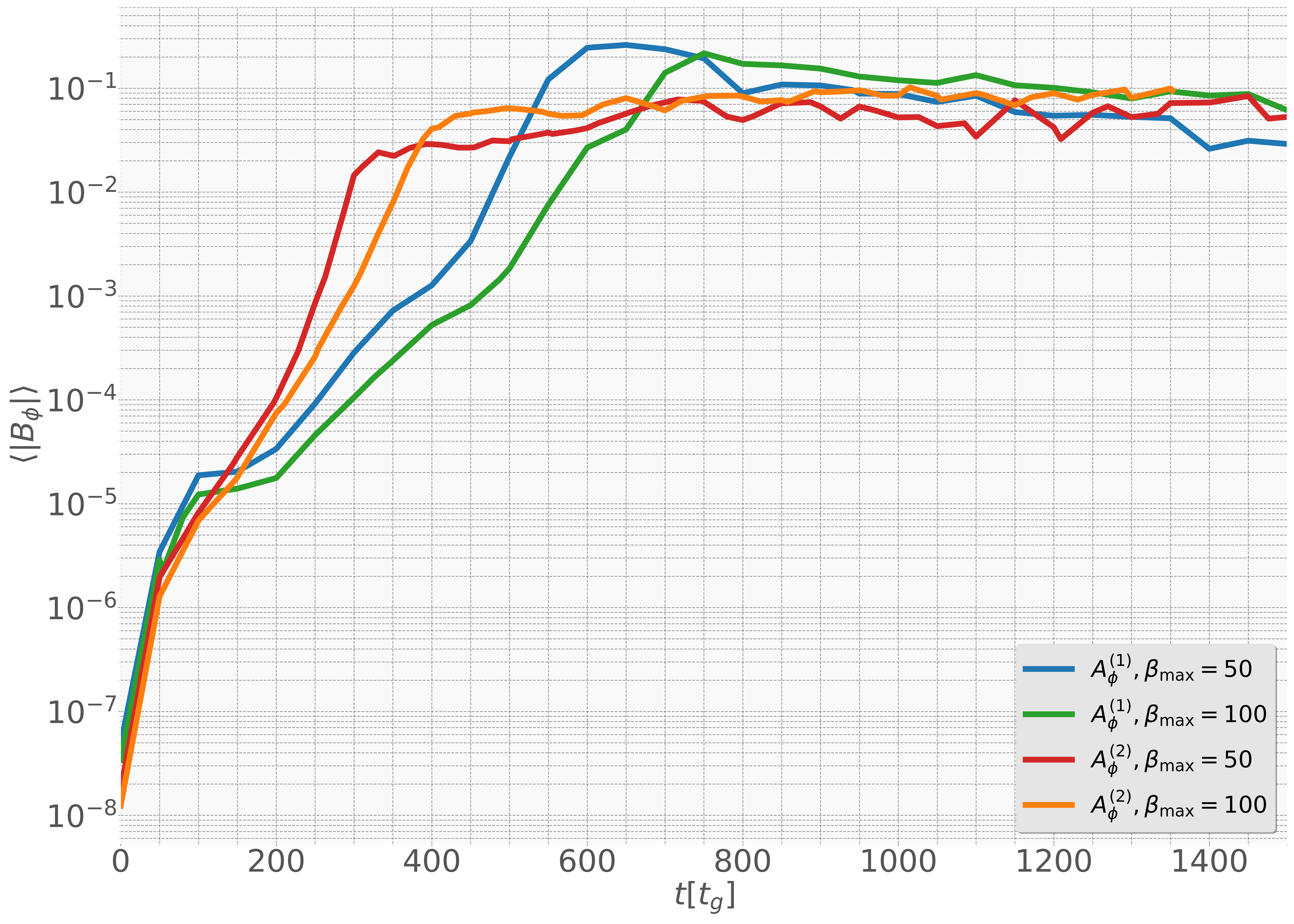}
    \caption{Time evolution of the (\textbf{left}) poloidal and (\textbf{right}) toroidal magnetic field strengths for all four simulation setups—$A_{\phi}^{(1)}$ and $A_{\phi}^{(2)}$ with initial plasma $\beta = 50$ and $100$.}
    \label{fig:magfield}
\end{figure}

Figure ~\ref{fig:magfield} shows the temporal evolution of the poloidal (left) and toroidal (right) magnetic field strengths for all four simulation setups—$A_{\phi}^{(1)}$ and $A_{\phi}^{(2)}$ with initial plasma $\beta = 50$ and $100$. The poloidal field in the $A_{\phi}^{(2)}$ cases starts at a lower magnitude compared to $A_{\phi}^{(1)}$, reflecting the initially weaker large-scale flux in this configuration. However, it grows steadily over time due to advection and accumulation of magnetic flux toward the black hole, eventually approaching comparable values to the $A_{\phi}^{(1)}$ runs. The toroidal field evolution is broadly similar across all four cases, indicating that the differential rotation of the disk efficiently winds up the magnetic field in all setups, largely independent of the initial vector potential. These trends suggest that while $A_{\phi}^{(1)}$ promotes an early onset of jet launching due to its initially strong poloidal flux, $A_{\phi}^{(2)}$ develops comparable poloidal strength more gradually, allowing a more disk-mediated route to MAD states.

\begin{figure}[H]
    \includegraphics[width=0.48\linewidth]{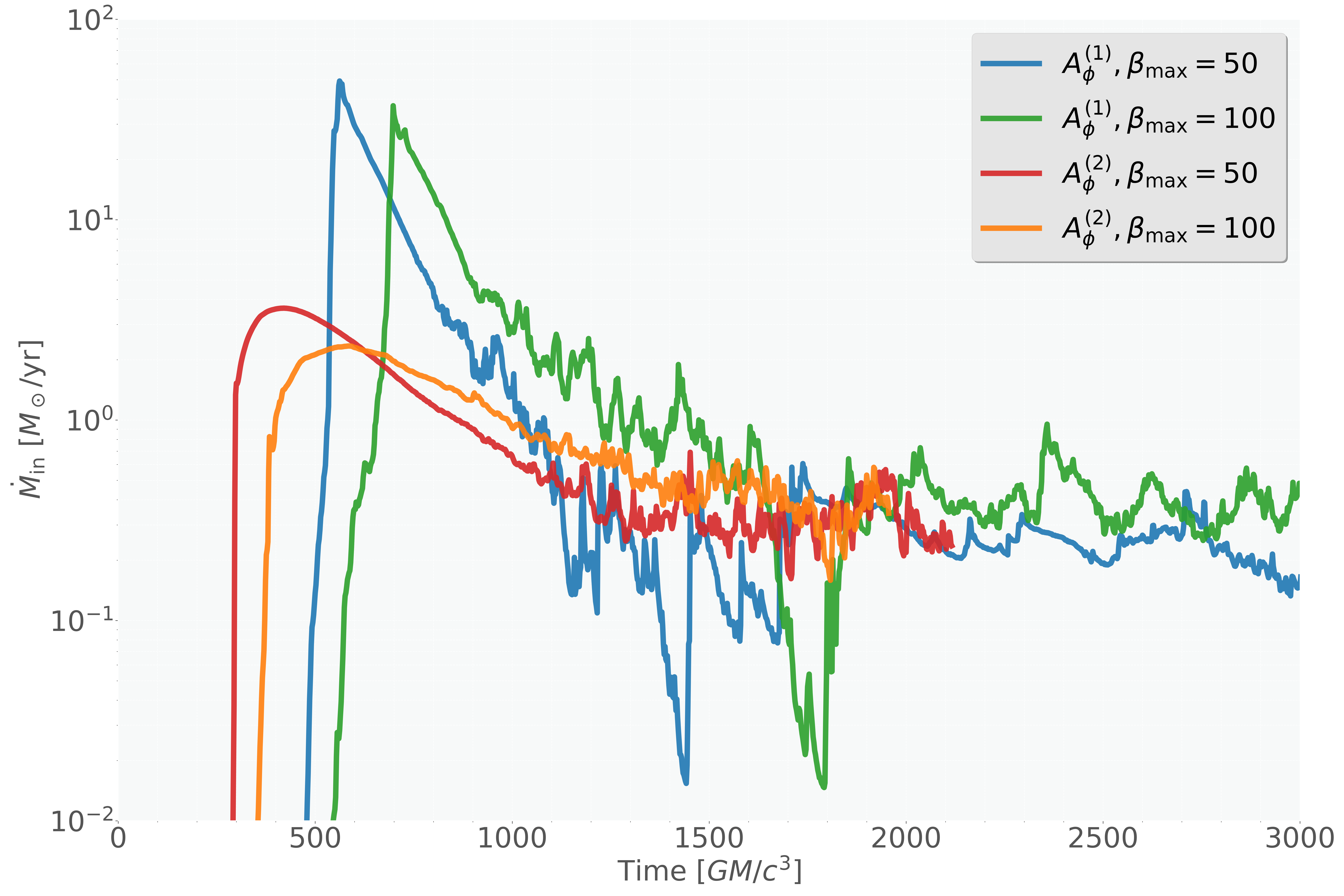}
    \includegraphics[width=0.48\linewidth]{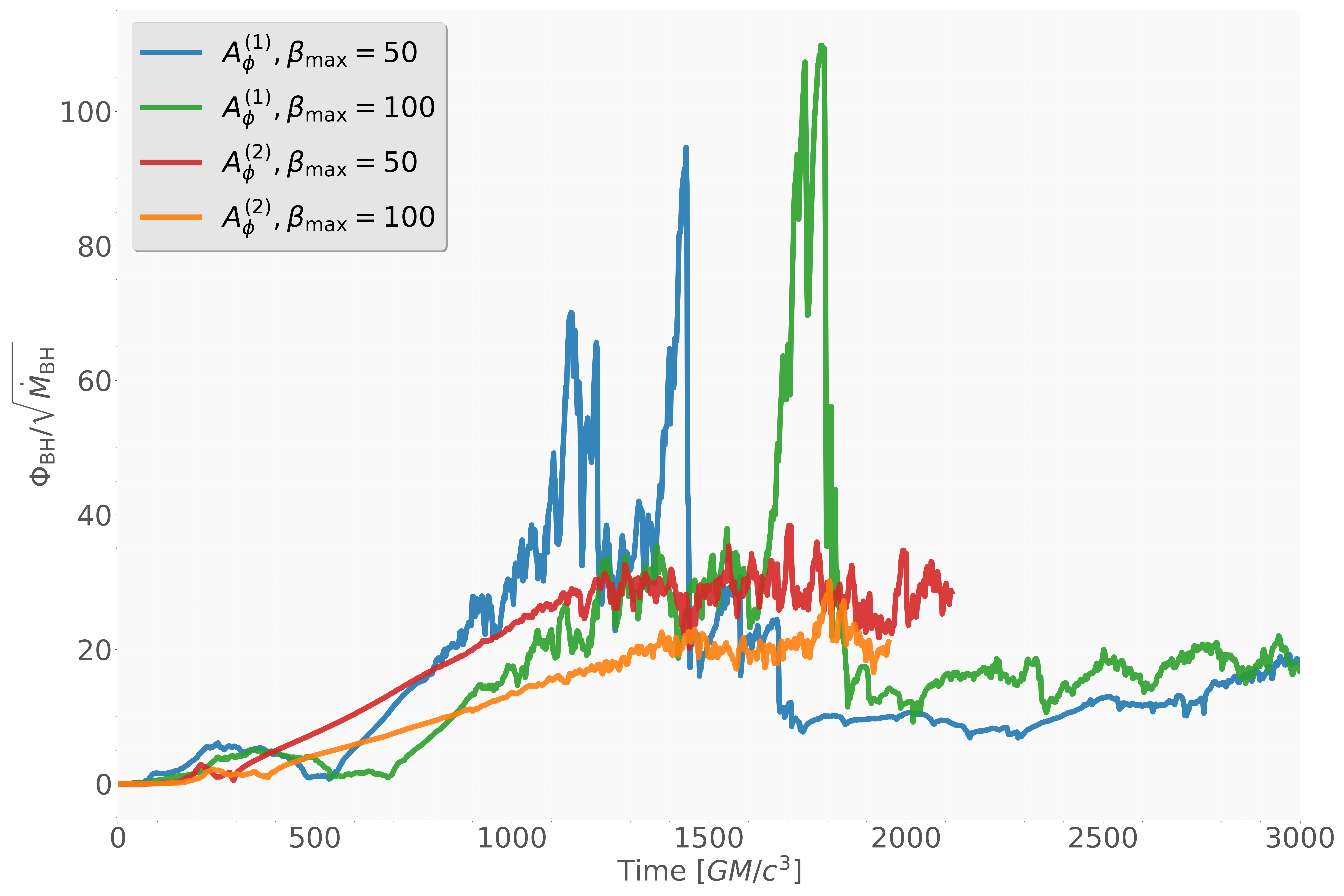}
    \caption{Mass accretion rate through the inner boundary (\textbf{left}) and normalized mass flux (\textbf{right}) of the system for vector potentials $A_{\phi}^{(1)}$ and $A_{\phi}^{(2)}$ at different plasma $\beta$.}
    \label{fig:mdot-phi}
\end{figure}

Figure~\ref{fig:mdot-phi} presents the time evolution of the mass accretion rate through the inner boundary and the normalized magnetic flux for all four simulation setups, where the latter is computed as 
\begin{align}
    \Phi_{\mathrm{BH}} = \frac{1}{2\sqrt{\dot M}}\int_\theta \int_\phi |B^r (r_H, t) | \, dA_{\theta\phi}. \label{magfluxnorm} 
\end{align}
The mass accretion rate drops more rapidly in the $A_{\phi}^{(1)}$ cases compared to $A_{\phi}^{(2)}$, consistent with the faster depletion of the torus in $A_{\phi}^{(1)}$ due to its initially stronger magnetic field and higher magnetization. The $A_{\phi}^{(2)}$ runs show a more gradual decrease in $\dot{M}$, reflecting the slower, disk-mediated evolution of the system. Correspondingly, the horizon-threading magnetic flux steadily increases in the $A_{\phi}^{(2)}$ configurations as the large-scale poloidal field is advected inward.

Previous studies have shown that the $A_{\phi}^{(1)}$ configuration reaches a steady state at $t \gtrsim 5 \times 10^{3}\,t_{g}$, with a typical accretion rate of $\dot{M} \sim 10^{-2}$ \cite{James2022}. In contrast, the $A_{\phi}^{(2)}$ configuration attains a quasi-steady state only at much later times, $t \gtrsim 10^{4}\,r_{g}/c$, with a higher accretion rate of $\dot{M} \sim 10^{0}$ \cite{Chatterjee2022}. In our simulations, the evolution is still in the early stages, and the system has not yet reached a fully developed quasi-steady state. At the resolution employed in our models, they display indications of approaching a quasi-steady accretion state during the evolution achieved to date.

\begin{figure}[H]
    \includegraphics[width=0.8\linewidth]{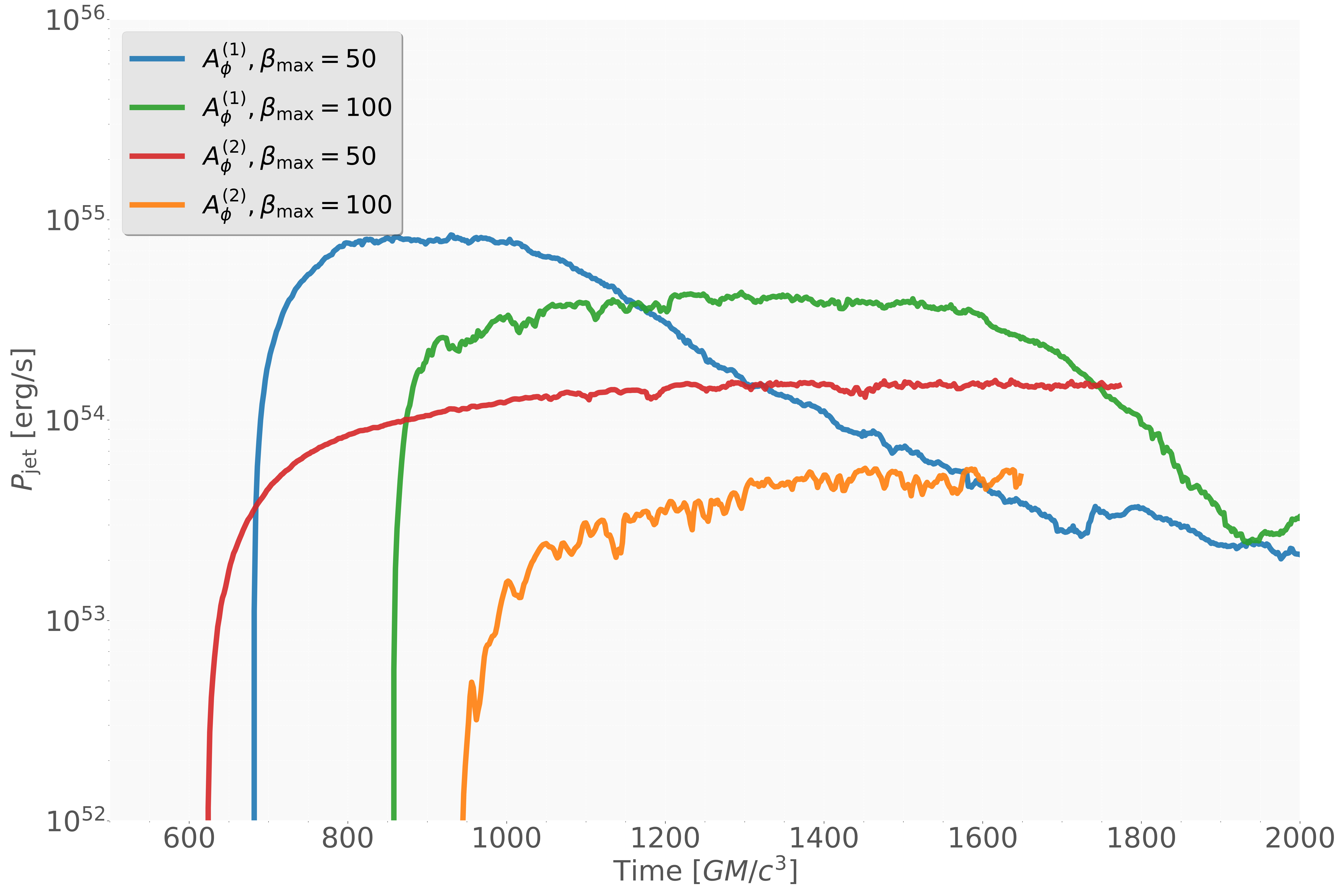}
    \caption{ Time evolution of the jet power $P_\mathrm{jet}$ for the four simulation setups: $A_{\phi}^{(1)}$ and $A_{\phi}^{(2)}$ with initial plasma $\beta = 50$ and $100$.}
    \label{fig:jetpower}
\end{figure}

Figure~\ref{fig:jetpower} shows the evolution of jet power for all four setups, calculated from the radial energy flux \cite{McKinney2004} as 
\begin{align}
    \dot E = \int_0^{2\pi } \int _0^\pi d\theta d\phi \sqrt{-g} F_E, \label{radialflux}
\end{align}where $F_E \equiv -T^r_t$. The $A_{\phi}^{(1)}$ configuration at $\beta = 50$ experiences a gradual drop in $P_\mathrm{jet}$, suggesting that this initial field geometry cannot efficiently sustain jets under stronger magnetization. This decrease in jet power can be attributed to the very strong initial poloidal flux, which leads to intermittent magnetic reconnection and disruption of the inner jet region. The highly magnetized plasma in this case is less stable, reducing the efficiency of energy extraction from the black hole and temporarily lowering the jet power.
Other models, $A_{\phi}^{(1)}$ at $\beta = 100$ and $A_{\phi}^{(2)}$ at both $\beta = 50$ and $100$ show gradual growth or nearly constant jet power, reflecting more stable magnetic field configurations and sustained \mbox{jet launching}.

\section{Conclusions and Discussion}\label{sec.4}

The differences between the two vector-potential configurations primarily arise from the scale, geometry, and net poloidal flux of the initial magnetic field. The configuration $A_{\phi}^{(1)}$ contains a larger-scale, coherent poloidal loop with strong initial flux near the inner disk, which enhances the MRI growth rate, increases angular momentum transport via Maxwell stresses, and leads to rapid mass accretion. This strong initial poloidal flux also quickly threads the black hole, powering an early, intense jet. However, in cases with very strong initial magnetization (e.g., $\beta = 50$), the jet power can drop gradually, indicating that this configuration is less stable under strong magnetic fields.

By contrast, $A_{\phi}^{(2)}$ starts with a weaker, more disk-centered poloidal field but with a smoother, large-scale topology. As the simulation evolves, magnetic flux is gradually advected inward, leading to steady growth of poloidal strength and a more coherent, ordered field throughout the disk and polar regions. This promotes a gradual buildup toward a MAD state, smoother jet boundaries, and a broader, well-collimated polar funnel. The torus also evolves more slowly, retaining more mass at comparable times, and the resulting jets are stable or steadily increasing in power across both $\beta = 50$ and $100$ cases.

It is important to emphasize that the primary goal of this study is not to perform long-term, fully relaxed GRMHD simulations or to reach a final quasi-steady state. Instead, our aim is to explore how different choices of initial vector potentials influence the early growth of magnetic flux, the evolution of the torus, and the onset of jet launching. This comparative approach could help in  the choice of vector potentials for magnetic field seeding, depending on the specific physical scenario of interest
such as GRB central engines, jet dynamics, or black hole growth—in future high-resolution, long-duration simulations. Our study focuses on this early phase, which plays a critical role in determining when and how a system approaches the SANE or MAD state.

Overall, these behaviors highlight how the choice of initial vector potential controls the early accretion and jet evolution. $A_{\phi}^{(1)}$ is suitable for studies focused on rapid jet launching and high mass-loss events, whereas $A_{\phi}^{(2)}$ provides a more realistic, disk-mediated path to MAD, allowing the study of gradual jet collimation and ordered magnetic field evolution. The complementary nature of these setups emphasizes the importance of initial field geometry in shaping both accretion dynamics and jet properties.

Recent studies \citep{Salas2024, James2022}, along with numerous works on MAD/SANE accretion \citep{Tchekhovskoy2011-4, Liska2020} \& references therein), have extensively examined the long-term evolution of black hole accretion flows and jet formation. Collectively, these investigations have established the canonical picture of MAD and SANE states and clarified how their properties depend on the accumulation of large-scale magnetic flux. While $A_{\phi}^{(2)}$ can be somewhat more computationally expensive than $A_{\phi}^{(1)}$, it provides a more physically informative evolution, producing a coherent, ordered magnetic field and smoother jet structures. Our simulations, in contrast, focus on the very early stages of evolution, approaching a quasi-steady state showing the difference in seeding of the magnetic field due to different vector potentials.

In summary, our results demonstrate that the initial vector potential is not merely a technical choice but a fundamental determinant of magnetic flux accumulation, torus dynamics, and jet launching efficiency. By highlighting how initial magnetic topology influences the early evolution of accretion flows and the timescale for MAD development, this work underscores the need to carefully select and characterize initial magnetic configurations in GRMHD studies. These findings offer a framework for interpreting early jet variability and magnetic flux growth in accreting black hole systems and provide a foundation for future investigations linking numerical models with observations of magnetically driven outflows. 

\vspace{6pt}

\authorcontributions{Software, Ishika Palit; 
Resources, Hsiang-Yi Karen Yang; 
Writing—original draft preparation, Ishika Palit; 
Writing—review and editing, Miles Angelo Sodejana and Hsiang-Yi Karen Yang; 
Visualization, Ishika Palit and Miles Angelo Sodejana; 
Supervision, Ishika Palit and Hsiang-Yi Karen Yang; 
Project administration, Hsiang-Yi Karen Yang. 
All authors have read and agreed to the published version of the manuscript.}

\funding{This research received no external funding }

\dataavailability{The simulation data used in this study are stored on the CICA computing cluster of the Institute of Astronomy, National Tsing Hua University (https://github.com/nthu-ioa/cluster) and are available from the corresponding author upon reasonable request. The HARM code employed for the simulations is publicly available and cited in the References.
} 

\acknowledgments {The simulations were performed using computational resources provided by the Center for Informatics and Computation in Astronomy (CICA) at National Tsing Hua University, funded by the Ministry of Education (MOE), Taiwan. I.P. acknowledges support from the National Science and Technology Council (NSTC), Taiwan, under Grant No. 112-2811-M-007-055-MY3. M.A.P.S. acknowledges the National Center for Theoretical Sciences—Theoretical and Computational Astrophysics (NCTS-TCA) Summer Student Program 2025 for providing the opportunity to conduct this research. H.-Y.K.Y. acknowledges support from the NSTC of Taiwan (NSTC 112-2628-M-007-003-MY3; NSTC 114-2112-M-007-032-MY3) and from the Yushan Scholar Program of the Ministry of Education (MOE) of Taiwan (MOE-108-YSFMS-0002-003-P1).} 

\conflictsofinterest{The authors declare no conflicts of interest.} 

\begin{adjustwidth}{-\extralength}{0cm}

\reftitle{References}



\begin{thebibliography}{999}

\bibitem[McKinney(2013)]{McKinney2013}
McKinney, J.C.; Tchekhovskoy, A.; Blandford, R.D.  Alignment of Magnetized Accretion Disks and
Relativistic Jets with Spinning Black Holes. {\em Science}
{\bf 2013}, {\em 339}, 49.
\url{https://doi.org/10.1126/science.1230811}.

\bibitem[Bardeen(1975)]{bardeen1975}
Bardeen, J.M.; Petterson, J.A.  The
Lense-Thirring Effect and Accretion Disks around Kerr
Black Holes. {\em Astrophys. J. Lett.} {\bf 1975}, {\em 195}, L65.
\url{https://doi.org/10.1086/181711}.

\bibitem[Lodato(2006)]{lodato2006}
Lodato, G.; Pringle, J.E.  The evolution of
misaligned accretion discs and spinning black holes.
{\em Mon. Not. R. Astron. Soc.} {\bf 2006}, {\em 368}, 1196--1208.
\url{https://doi.org/10.1111/j.1365-2966.2006.10194.x}.

\bibitem[Tchekhovskoy(2011)]{Tchekhovskoy2011-4}
Tchekhovskoy, A.; Narayan, R.; McKinney, J.C. Efficient generation of jets from magnetically arrested accretion on a rapidly spinning black hole.
 {\em Mon. Not. R. Astron. Soc. Lett.} {\bf 2011}, {\em 418}, L79--L83.
\url{https://doi.org/10.1111/j.1745-3933.2011.01147.x}.

\bibitem[Chatterjee(2022)]{Chatterjee2022}
Chatterjee, K.; Narayan, R. Flux Eruption Events Drive Angular Momentum Transport in Magnetically Arrested Accretion Flows. {\em  Astrophys. J.} {\bf 2022}, {\em 941}, 30.
\url{https://doi.org/10.3847/1538-4357/ac9d97}.

\bibitem[Salas(2024)]{Salas2024}
Salas, L.D.S.; Musoke, G.; Chatterjee, K.; Markoff, S.B.; Porth, O.; Liska, M.T.P.; Ripperda, B. Resolution analysis of magnetically arrested disc simulations. {\em Mon. Not. R. Astron. Soc.} {\bf 2024}, {\em 533}, 254--267.
\url{https://doi.org/10.1093/mnras/stae1834}.


\bibitem[EHT(2024)]{eht2024}  Akiyama, K. et al. [Event Horizon Telescope Collaboration].
 The persistent shadow of the supermassive black hole of M87. {\em Astron. Astrophys.} {\bf 2024}, {\em 681}, 63.
\url{https://doi.org/10.1051/0004-6361/202347932}.

\bibitem[Chael(2019)]{chael2019} Chael, A.; Narayan, R.; Johnson, M.D. Two-temperature, Magnetically Arrested Disc simulations of the jet from the supermassive black hole in M87. {\em Mon. Not. R. Astron. Soc.} {\bf 2019},  {\em 486}, 2873--2895. 
\url{https://doi.org/10.1093/mnras/stz988}.

\bibitem[Chan(2025)]{chan2025} Chan, H.-S.; Chan, C. The 230 GHz Variability of Numerical Models of Sagittarius A*. II. The Physical Origins of the Variability.
{\em  Astrophys. J.} {\bf 2025}, {\em 985}, 164. 
\url{https://doi.org/10.3847/1538-4357/adc99f}.

\bibitem[Palit(2025)]{Palit2025}
Palit, I.; Dihingia, I.K.; Mizuno Y.; Yang H.Y.K. Obliquity of Black Hole Magnetosphere and its Impact on Accretion Dynamics. {\em{Astrophys. J.}} under review, 2025. 



\bibitem[Gammie(2003)]{Gammie2003}
Gammie, C.F.; McKinney, J.C.; Tóth, G. HARM: A Numerical Scheme for General Relativistic Magnetohydrodynamics. {\em Astrophys. J.} {\bf 2003}, {\em 589}, 444--457. \url{https://doi.org/10.1086/374594}.


\bibitem[Noble(2006)]{Noble2006}
Noble, S.C.; Gammie, C.F.; McKinney, J.C.; Del Zanna, L. Primitive Variable Solvers for Conservative General Relativistic Magnetohydrodynamics. {\em Astrophys. J.} {\bf 2006}, {\em 641}, 626--637. \url{https://doi.org/10.1086/500349}.

\bibitem[Sapountzis(2019)]{Sapountzis2019}
Sapountzis, K.; Janiuk, A. The MRI Imprint on the Short-GRB Jets.  {\em Astrophys. J.} {\bf 2019}, {\em 873}, 12--21. 
\url{https://doi.org/10.3847/1538-4357/ab0107}.

\bibitem[FM(1976)]{FM76}
Fishbone, L.G.; Moncrief, V. Relativistic fluid disks in orbit around Kerr black holes. {\em Astrophys. J.} {\bf 1976}, {\em 207}, 962--976. \url{https://doi.org/10.1086/154565}.



\bibitem[Igor(2003)]{Igor2003}
Igumenshchev, I.V.; Narayan, R.; Abramowicz, M.A. Three-dimensional Magnetohydrodynamic Simulations of Radiatively Inefficient Accretion Flows. {\em Astrophys. J.} {\bf 2003}, {\em 592}, 1042--1059. \url{https://doi.org/10.1086/375769}.

\bibitem[Narayan(2003)]{Narayan2003}
Narayan, R.; Igumenshchev, I.V.; Abramowicz, M.A. Magnetically Arrested Disk: An Energetically Efficient Accretion Flow. {\em Publ. Astron. Soc. Jpn.} {\bf 2003}, {\em 55}, L69--L72. \url{https://doi.org/10.1093/pasj/55.6.L69}.


\bibitem[James(2022)]{James2022}
James, B.; Janiuk, A.; Hossein Nouri, F. Modeling the Gamma-Ray Burst Jet Properties with 3D General Relativistic Simulations of Magnetically Arrested Accretion Flows. {\em Astrophys. J.} {\bf 2022}, {\em 935}, 176. \url{https://doi.org/10.3847/1538-4357/ac81b7}.

\bibitem[McKinney(2004)]{McKinney2004}
McKinney, J.C.; Gammie, C.F. A Measurement of the Electromagnetic Luminosity of a Kerr Black Hole. {\em Astrophys. J.} {\bf 2004}, {\em 611}, 977--995. \url{https://doi.org/10.1086/422244}.

\bibitem[Liska(2020)] {Liska2020} Liska ,M.; Tchekhovskoy, A.; Quataert, E. Large-scale poloidal magnetic field dynamo leads to powerful jets in GRMHD simulations of black hole accretion with toroidal field. \emph{Mon. Not. R. Astron. Soc.} \textbf{2020}, \emph{494}, 3656--3662. 
\url{https://doi.org/10.1093/mnras/staa955}.



\end{thebibliography}


\PublishersNote{}
\end{adjustwidth}
\end{document}